\documentclass[aps,pra,twocolumn,footinbib]{revtex4-1} 
\usepackage{amsmath}
\usepackage{graphicx}% Include figure files
\usepackage{dcolumn}% Align table columns on decimal point
\usepackage{bm}% bold math
\usepackage{braket}
\usepackage[usenames, dvipsnames]{color}
\definecolor{brown}{rgb}{0.6, 0.3, 0.06}

\usepackage{enumitem}
\usepackage{hyperref}
\hypersetup{
	colorlinks=true,
	linkcolor=blue,
	filecolor=blue,
	citecolor=blue,  
	urlcolor=black,
}

\newtheorem{theorem}{Theorem}

\newtheorem{lemma}[theorem]{Lemma}

\newenvironment{proof}[1][Proof]{\noindent\textbf{#1.} }{\ \rule{0.5em}{0.5em}}
\def\be{\begin{equation}}
	\def\ee{\end{equation}}
\def\ba{\begin{eqnarray}}
	\def\ea{\end{eqnarray}}

\usepackage{amsfonts}
\usepackage{subfigure}
\usepackage{bm}
\usepackage{xr}
\usepackage{mfirstuc}
\begin{document}
\date{\today}
\title{
Security proof framework for two-way Gaussian quantum key distribution protocols
}

\author{Quntao Zhuang$^{1,2}$}
\email{zhuangquntao@gmail.com}
\author{Zheshen Zhang$^{1,3}$}
\author{Norbert L\"utkenhaus$^{4,5}$}
\author{Jeffrey H. Shapiro$^1$}
\affiliation{$^1$Research Laboratory of Electronics, 
Massachusetts Institute of Technology, Cambridge, Massachusetts 02139, USA\\
$^2$Department of Physics, 
Massachusetts Institute of Technology, Cambridge, Massachusetts 02139, USA\\
$^3$Department of Materials Science and Engineering,
University of Arizona, Tucson, Arizona 85721, USA\\
$^4$Perimeter Institute for Theoretical Physics, 31 Caroline St N, Waterloo, Ontario N2L 2Y5, Canada\\
$^5$Institute for Quantum Computing and Department of Physics and Astronomy, University of Waterloo, Waterloo, Ontario N2L 3G1, Canada}
\date{\today}

\begin{abstract}
Two-way Gaussian protocols have the potential to increase quantum key distribution (QKD) protocols' secret-key rates by orders of magnitudes~[Phys.~Rev.~A {\bf 94}, 012322 (2016)]. Security proofs for two-way protocols, however, are underdeveloped at present. In this paper, we establish a security proof framework for the general coherent attack on two-way Gaussian protocols in the asymptotic regime. We first prove that coherent-attack security can be reduced to collective-attack security for all two-way QKD protocols.  Next, we identify two different constraints that each provide intrusion parameters  which bound an eavesdropper's coherent-attack information gain for any two-way Gaussian QKD protocol.  Finally, we apply our results to two such protocols.
\end{abstract} 
\maketitle

\section{Introduction}

The continuing improvement of classical computational power~\cite{classical_reference} and the emergence of quantum computers~\cite{chuang1998experimental,vandersypen2001experimental,castelvecchi2017quantum}, are increasing the likelihood that complexity-based classical cryptographic algorithms---such as Rivest-Shamir-Adleman encryption~\cite{rivest1978method} and elliptic-curve cryptography~\cite{koblitz1987elliptic,miller1985use}---will be broken. Two distinct approaches have emerged for countering this vulnerability: post-quantum cryptography~\cite{bernstein2009introduction}, which seeks new public-key cryptography algorithms that are immune to the threat posed by a quantum computer running Shor's algorithm~\cite{Shor_1997}; and quantum key distribution~\cite{Bennett20147} (QKD), which provides protocol security based on physical laws rather than computational complexity. 

In QKD, Alice and Bob establish a raw key by quantum-channel transmission and detection of photons. They use security testing and classical communication to quantify Eve's intrusion on the quantum channel. With these intrusion parameters they can place an upper bound on Eve's information gain. Then, they complete the QKD protocol by reconciling their raw keys, to eliminate errors, and distilling a final key via privacy amplification,  to ensure its unconditional protocol security.  

QKD's principal advantage is its provable protocol security.  Its ultimate benefit would be enabling Alice and Bob to transmit messages using one-time-pad encryption, which would afford them information-theoretic security for their communications.  QKD systems using the decoy-state BB84 or conventional continuous-variable (CV) protocols, however, have state-of-the-art secret key rates (SKRs)~\cite{tang2014measurement,lucamarini2013efficient,huang2015continuous} of $\sim$1\,Mbit/s at metropolitan-area distances, which is far below the Gbit/s rates needed for Internet-speed secure communications.  These systems' SKRs could be pushed to Gbit/s with massive combinations of space-division and wavelength-division multiplexing, but that approach comes with a major equipment burden in cost and complexity.  Recently, floodlight QKD (FL-QKD)~\cite{Quntao_2015,zhuang2017large,zhang2016floodlight,zhang2017floodlight} has been proposed as a means to realize Gbit/s SKRs at metropolitan-area distances over single-mode fiber (no space-division multiplexing), in a single-wavelength channel (no wavelength-division multiplexing), and without the need to develop any new technology.  It does so by encoding each transmitted symbol over multiple temporal modes, whereas decoy-state BB84 makes no use of multimode encoding and conventional CV-QKD requires single-mode encoding.  As a result, FL-QKD's SKR is less constrained by the PLOB bound~\cite{pirandola2017fundamental,footnote0}, which sets the ultimate limit on secret bits per mode, than those protocols.  That said, decoy-state BB84 and conventional CV-QKD have the advantage of being one-way (OW) protocols, whereas FL-QKD is a two-way (TW) protocol, so that the former have much stronger security guarantees---e.g., decoy-state BB84 has coherent-attack security with finite-key analysis---while the latter's security to date is only against the frequency-domain collective attack in the asymptotic regime~\cite{Quntao_2015}.  On the other hand, unlike other TW-QKD protocols~\cite{Pirandola_2008,Ping_Pong,two_way_no_loss,Han_2014,Weedbrook_2014,generalization_no_loss,Zhang_2014,Ottaviani_2015,Ottaviani_2016}, FL-QKD uses an optical amplifier in Bob's terminal to overcome the Bob-to-Alice channel's loss, making FL-QKD's channel loss equivalent to that of OW-QKD protocols.  

The limited nature of FL-QKD's security proof is characteristic of the situation for other TW-QKD protocol's~\cite{Quntao_2015,zhuang2017large,zhang2016floodlight,zhang2017floodlight,Ping_Pong,Pirandola_2008,two_way_no_loss,Han_2014,Zhang_2014,generalization_no_loss,Weedbrook_2014,Ottaviani_2015,Ottaviani_2016}.  In part, this is because proof techniques for OW-QKD~\cite{lo1999unconditional,shor2000simple,de_Finetti1,de_Finetti2,Leverrier_2013} do not readily cope with simultaneous attacks on both the Alice-to-Bob and Bob-to-Alice channels of a TW-QKD protocol.  Thus, for a long time only special attacks~\cite{Han_2014,Pirandola_2008,Weedbrook_2014,Zhang_2014,Ottaviani_2015,Ottaviani_2016,Sun_2012}, or general attacks in the absence of loss and/or noise~\cite{Ping_Pong,two_way_no_loss,generalization_no_loss}, have been considered for TW-QKD.  

At this point it should be clear that a coherent-attack security proof, with finite-key analysis~\cite{renner2008security}, for FL-QKD would be an enormous step toward its widespread employment, given that protocol's potential for Gbit/s SKRs.  More generally, such security proofs for other TW-QKD protocols could also be valuable.  In this paper, we take a first step in that direction by establishing TW-QKD's coherent-attack security in the asymptotic regime wherein there are an infinite number of channel uses.  Because we want to include protocols like FL-QKD that encode over multiple modes, we cannot entirely rely on raw-key post-processing by Alice and Bob that permutes or otherwise independently manipulates measurements made on individual optical modes. Moreover, at high modulation rates, manipulation of each encoded symbol is difficult, even if single-mode encoding is employed. Hence, de~Finetti theorem arguments~\cite{de_Finetti1,de_Finetti2} are not directly applicable for the reduction from coherent attack to collective attack on a mode-by-mode basis; Instead, it reduces the general coherent attack to a block-wise coherent attack, in which Eve performs the same but arbitrary operations on blocks of modes that may comprise one or more symbols, i.e., one or more of Alice and Bob's channel uses. 

The asymptotic SKR for the block-wise coherent attack is given by the block-wise Devetak-Winter formula~\cite{devetak2005distillation,RMP_security}, which reduces the problem to bounding the information leaked to Eve during the raw-key generation process, as constrained by intrusion-parameter estimates made by Alice and Bob using local operations and classical communication (LOCC)~\cite{nielsen1999conditions}. 
Note that in a block-wise coherent attack~\cite{renner2008security}, Eve can entangle signals sent in different channel uses within a block, which makes bounding Eve's information gain a difficult multi-letter maximization involving Eve's operations over multiple channel uses.  A crucial task for us is therefore reducing the maximization to a single-letter (single channel-use) form from which computing an upper bound on Eve's information gain is tractable.  

We will use the recently developed noisy-entanglement-assisted classical capacity formula~\cite{Zhuang_2016_cl,Quntao_2015} to resolve the preceding single-letterization dilemma for Gaussian TW-QKD protocols, i.e., protocols that employ Gaussian-state sources and Gaussian operations~\cite{Weedbrook_2012}. We consider two different constraints---the pair-wise sum constraints and the permutation-invariant sum constraints---that provide intrusion parameters which suffice to establish tractable  (single-letter) upper bounds on Eve's coherent-attack information gain.   

The remainder of the paper is organized as follows. In Sec.~\ref{intro_TWQKD}, we introduce Gaussian TW-QKD protocols. In Sec.~\ref{bound_TWQKD}, we prove that the capacity formula in Ref.~\cite{Zhuang_2016_cl} provides an upper bound on Eve's information gain from her most general coherent attack.  There, we also describe the pair-wise sum constraints and permutation-invariant sum constraints that lead to efficiently calculable bounds on Eve's coherent-attack information gain, and we show that the resulting upper bound can be achieved by a collective attack. In Sec.~\ref{SKR_various}, we evaluate the secret-key efficiencies (in bits/symbol) of two Gaussian TW-QKD protocols:  the two-mode squeezed-vacuum (TMSV) protocol from Refs.~\cite{Pirandola_2008,Weedbrook_2014,Ottaviani_2015,Ottaviani_2016} and FL-QKD.  We conclude, in Sec.~\ref{discussion}, with a summary and some discussion.  

\section{Gaussian two-way protocols}
\label{intro_TWQKD}
\begin{figure}
\includegraphics[width=0.35\textwidth]
{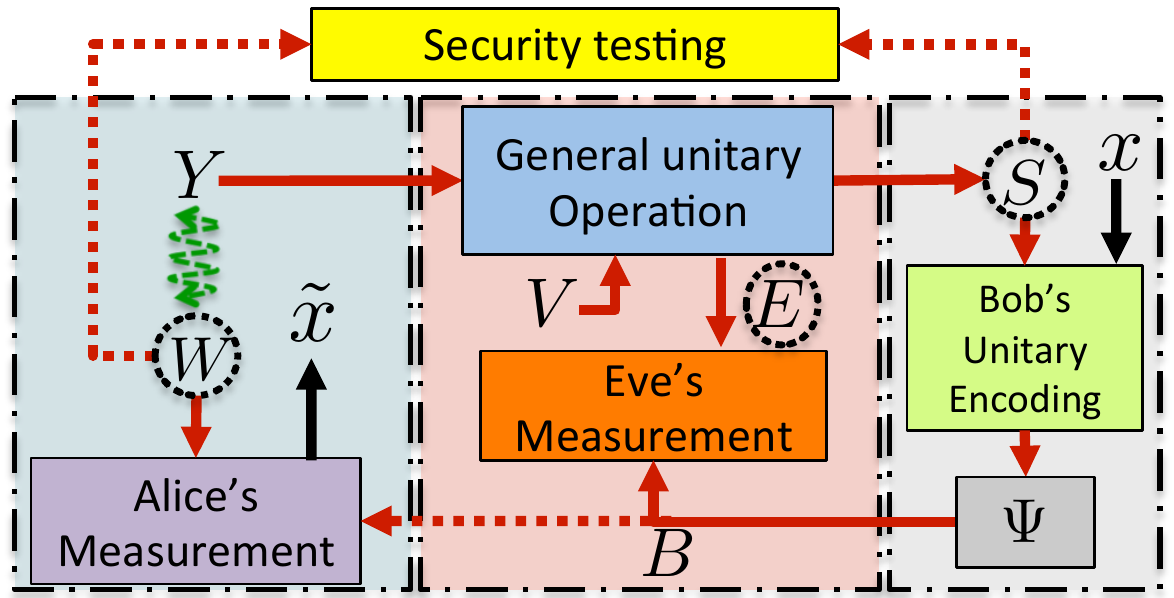}
\caption{General schematic for a Gaussian TW-QKD protocol.}
\label{scheme_QKD}
\end{figure}

Figure~\ref{scheme_QKD} shows a general schematic for how a Gaussian TW-QKD protocol generates raw key~\cite{footnote1}.  First, Alice prepares a signal-reference mode pair $(Y,W)$ in a maximally entangled bipartite Gaussian pure state, i.e., the two-mode squeezed vacuum (TMSV), with average photon number $N_S$ and Wigner covariance matrix
\begin{equation}
{\boldsymbol \Lambda}_{YW} =  \frac{1}{4}\left[\begin{array}{cc}
{\bf A}_Y & {\bf C}_{YW} \\
{\bf C}_{YW} & {\bf A}_W \end{array}\right],
\label{cov_TMSV}
\end{equation}
where ${\bf A}_Y ={\bf A}_W = (2N_S+1){\bf I}_2$, and ${\bf C}_{YW} = 2C_S\,{\bf Z}_2$, with ${\bf I}_2={\rm Diag}[1,1]$, ${\bf Z}_2={\rm Diag}[1,-1]$, and $C_S=\sqrt{N_S(N_S+1)}$.
As explained below, Alice will use part of $W$ for security testing~\cite{footnote2}, but Eve is unable to access any of $W$. Next, Alice sends $Y$ to Bob through a quantum channel that Eve controls.  In general, Eve performs a unitary operation on $Y$ and some pure-state modes $V$ of her own, producing her ancilla $E$, which can be multi-mode, and the single-mode signal $S$ that she transmits to Bob.  Because Eve can mount a coherent attack, her unitary operation can act on the entire sequence of signals that Alice transmits during a QKD session.

Bob takes a small portion of the $S$ mode he receives from Eve and uses it for security testing. After that, he encodes a random symbol $X$ on the remainder of the $S$ mode to produce his encoded mode $S'$.  He does so, when $X=x$, by means of a unitary $U_x$ that imparts a deterministic, complex-valued displacement $d_x$, and a deterministic phase shift $\theta_x$.  The $S'$ mode's photon annihilation operator is  therefore
$\hat{a}_S^{\prime}=e^{i\theta_X}\hat{a}_S+d_X,$
where $\hat{a}_S$ is the $S$ mode's annihilation operator, and we have not accounted for the small portion of $S$ that is consumed by Bob's security testing.  

We will assume that $d_X$ is a zero-mean, circulo-complex, Gaussian random variable. Thus, if no phase encoding is applied, then $S'$ will be in a thermal state with average photon number 
\be
\braket{\hat{a}_S^{\prime\dagger} \hat{a}_S^\prime}=\int\!{\rm d}x\,p_X(x) \braket{\hat{a}_S^{\prime\dagger} \hat{a}_S^\prime}_x=\braket{\hat{a}_S^\dagger \hat{a}_S}+E_X,
\label{ex_power}
\ee
where $\langle \cdot\rangle_x$ denotes averaging conditioned on $X=x$, and $E_X \equiv \int\!{\rm d}x\,p_X(x) |d_x|^2$ with $p_X(x)$ being $X$'s probability density function.  
Our security analysis will presume encoding symmetry, i.e., that the $S'$ mode's unconditional state has zero mean, which is guaranteed by assuming that 
$\int\!{\rm d}x\, p_X(x) e^{i\theta_x}=0$.  Note that the $S'$ mode's average photon number is unaffected by the phase shift $\theta_X$, making Eq.~(\ref{ex_power}) applicable when Bob encodes in both displacement and phase.  

The preceding encoding scheme, while not the most general, includes the random-displacement encoding employed in Ref.~\cite{Pirandola_2008}, and the phase encoding used in FL-QKD~\cite{Quntao_2015,zhuang2017large,zhang2016floodlight,zhang2017floodlight}. Here we note that the average state of $(S^\prime,W)$ is, in general, non-Gaussian owing to the action of Eve's unitary operation and/or Bob's phase modulation.  

After completing his encoding, Bob sends $S$' through a quantum channel, $\Psi$, within his terminal that models the characterizable part of the Bob-to-Alice return path that is not controlled by Eve.  This channel  produces an output mode $B$ that Bob sends to Alice through a quantum channel that is under Eve's control.  Alice makes a joint measurement of the mode she receives and $W$ to obtain her raw-key symbol $\tilde{x}$ that results from Bob's having sent $x$, with the nature of Alice's measurement depending on Bob's choice of his encoding operation $U_X$.  Alice and Bob also perform security testing, which is an LOCC parameter-estimation scheme based on Alice's measuring part of $W$ and Bob's measuring part of $S$.  That scheme allows them to evaluate some bipartite functions of the joint state $\rho_{SW}$ that constitute intrusion parameters which they use to compute an upper bound on Eve's information gain.  

Alice and Bob distill their secret key by the following two-step procedure.  Starting from his  transmitted-symbol sequence and Alice's raw-key sequence, Bob performs the key-map operation~\cite{renner2008security,keymap} and sends error-correction information to Alice on an authenticated classical channel.  At that point, Alice and Bob share a common key, but its security is not assured because of the information Eve has gained.  Thus they use their upper bound on Eve's information gain to determine and perform a sufficient amount of privacy amplification to ensure their final key's security.

To complete our explanation of Fig.~\ref{scheme_QKD}, we conclude this section with some remarks about $\Psi$, the quantum channel within Bob's terminal.  For single-mode encoding, we take $\Psi$ to be a single-mode Gaussian channel with no excess noise, which can be represented as a unitary operation on the encoded signal mode $S'$ and a vacuum-state environment mode $N$ that produces the return mode $B$ and a transformed environment mode $N^\prime$.  
If multi-mode encoding over $M_E > 1$ modes is employed, the channel internal to Bob's terminal is $\Psi^{\otimes M_E}$, which applies, e.g., to FL-QKD with $\Psi$ being a quantum-limited amplifier channel, $\mathcal{A}_{G_B}^{0}$, whose output modes $B$ and $N'$ are characterized by
$\hat{a}_B=\sqrt{G_B}\,  \hat{a}_S^\prime+\sqrt{G_B-1}\, \hat{a}_N^\dagger,$
and $\hat{a}_N' = \sqrt{G_B-1}\,\hat{a}_S^{\prime\dagger} + \sqrt{G_B}\,\hat{a}_N,$
where $G_B\ge 1$.
The $\Psi$ channel can also model loss in Bob's terminal by means of a pure-loss channel, $\mathcal{L}_{\eta_B}^{0}$, whose output modes satisfy
$\hat{a}_B=\sqrt{\eta_B}\,\hat{a}_S^\prime+\sqrt{1-\eta_B}\,\hat{a}_N,$
and $\hat{a}_N' = \sqrt{1-\eta_B}\,\hat{a}_S' - \sqrt{\eta_B}\,\hat{a}_N$, 
where $0\le \eta_B\le1$. For completeness, we will also consider the complement of $\mathcal{A}_{G_B}^{0}$---the contravariant quantum-limited amplifier channel $\tilde{\mathcal{A}}_{G_B}^{0}$---whose outputs obey
$\hat{a}_B=\sqrt{G_B-1}\,\hat{a}_S^{\prime\dagger} + \sqrt{G_B}\,\hat{a}_N  $
and $\hat{a}_N' = \sqrt{G_B}\,\hat{a}_S^\prime + \sqrt{G_B-1}\,\hat{a}_N^\dagger$.  From these input-output relations the $B$ mode's average photon number, $\langle \hat{a}_B^\dagger\hat{a}_B\rangle$, can be found to be
\be
N_B=
\left\{\begin{array}{cc}
G_B(\braket{\hat{a}_S^{\prime\dagger} \hat{a}_S^\prime}+E_X)+G_B-1, \mbox{ for } \mathcal{A}_{G_B}^{0}, \\[.05in]
\eta_B(\braket{\hat{a}_S^{\prime\dagger} \hat{a}_S^\prime}+E_X),  \mbox{ for } \mathcal{L}_{\eta_B}^{0},\\[.05in]
(G_B-1)(\braket{\hat{a}_S^{\prime\dagger} \hat{a}_S^\prime}+E_X +1),  \mbox{ for } \tilde{\mathcal{A}}_{G_B}^{0}.
\end{array}\right.
\label{Nb}
\ee
In deriving an upper bound on Eve's information gain we can (and will) assume that Eve collects \emph{all} the $B$ modes, because Bob performs the key-map operation.  

As a final note on $\Psi$, we point out that no loss of generality is entailed by our assumption that $\hat{a}_N$ is in its vacuum state.  This is because Eve gains less information when $\hat{a}_N$ is in a thermal state than when that mode is in its vacuum state, as is easily demonstrated by a channel decomposition~\cite{caruso2006one,garcia2012majorization} argument. In particular, let an $N_0$ superscript on our channel models' symbols denote the average photon number of that channel's thermal-state environment.  The thermal-environment amplifier channel can be expressed as 
\be 
\mathcal{A}_{G_B}^{N_0}=\mathcal{L}_{1/G'}^{N_0^\prime} \circ \mathcal{A}_{G^\prime}^{0}\circ \mathcal{A}_{G_B}^{0},
\ee 
where $G^\prime=\sqrt{1+N_0^\prime}/\sqrt{1+N_0^\prime-N_0(G_B^2-1)}>1$ with $N_0^\prime>N_0(G_B^2-1)$.  Likewise, the thermal-environment loss channel can be written as
\be
\mathcal{L}_{\eta_B}^{N_0}=\mathcal{A}_{1/\eta^\prime}^{0} \circ \mathcal{L}_{\eta^\prime}^{0} \circ \mathcal{L}_{\eta_B}^{0},
\ee
where $\eta^\prime=1/{\sqrt{1+N_0(1-\eta_B^2)}}<1$. (A similar relation holds for the complementary channel, but we shall omit it).  The data-processing inequality~\cite{nielsen2010quantum} now guarantees that the upper bound on Eve's information gain for the $N_0= 0$ version of each of our Gaussian $\Psi$ channels is also an upper bound on the information Eve gains from the $N_0 >0$ version of that Gaussian channel.

\section{Bounding Eve's information gain}
\label{bound_TWQKD}

For protocols that encode multiple modes per symbol, post-processing cannot independently manipulate each mode within a raw-key symbol.  In particular, permutation of the raw keys only permutes multiple-mode blocks.
Hence, the de~Finetti theorem~\cite{de_Finetti1,de_Finetti2} can be used to reduce the coherent attack to a block-wise coherent attack, but not to reduce it further to a single-mode attack. 
In a block-wise coherent attack, Eve performs the same arbitrary attack on each size $M_B \gg 1$ symbol block of Alice and Bob's transmissions. To accomplish the reduction, we will make use of the tools from Ref.~\cite{Zhuang_2016_cl}.

Consider a multiple-block QKD session in which Alice and Bob spend some small amount of pre-shared key to randomly discard some of the size-$M_B$ blocks, during post-processing of their raw keys. By de~Finetti theorem, we only need to bound Eve's information gain by analyzing, see Sec.~\ref{block_bound}, the block-wise coherent attack. Note that because $M_B\gg1$, the amount of key consumption for determining the blocks being discarded is very small compared to the keys being generated. It is worth emphasizing that there is hope that the reduction from the general coherent attack to the block-wise coherent attack may be accomplished without relying on de~Finetti argument, see Sec.~\ref{block_reduce}.

Alice and Bob's secret-key efficiency (SKE), in bits/symbol, for an asymptotic-regime block-coherent attack is given by the Devetak-Winter formula~\cite{devetak2005distillation,leverrier2009unconditional,RMP_security}
\be 
{\rm SKE} = \max(\beta I_{AB}-M_E \chi_E,0).
\label{DW_rate}
\ee 
Here: $I_{AB}$ is the Shannon information (in bits/symbol) between Bob's key map $\{X\}$ and Alice's measurement data $\{\tilde{X}\}$;  $\beta$ is Alice and Bob's reconciliation efficiency; $\chi_E$ is Eve's bits/mode Holevo-information gain; and $M_E$ is the number of modes per encoded symbol. Alice and Bob can calculate $I_{AB}$ from error-probability measurements, and they know the efficiency of their error-correction procedure, but they need to maximize $\chi_E$ over all block-wise coherent attacks that are consistent with their security-testing results.  That maximization---obtaining an upper bound on $\chi_E$---is therefore the heart of the asymptotic-regime security proof for TW-QKD protocols. In what follows, we show that the structure of TW-QKD protocols leads to an additive upper bound on $\chi_E$ that, in turn, results in an SKE lower bound.

\begin{figure}
\includegraphics[width=0.35\textwidth]{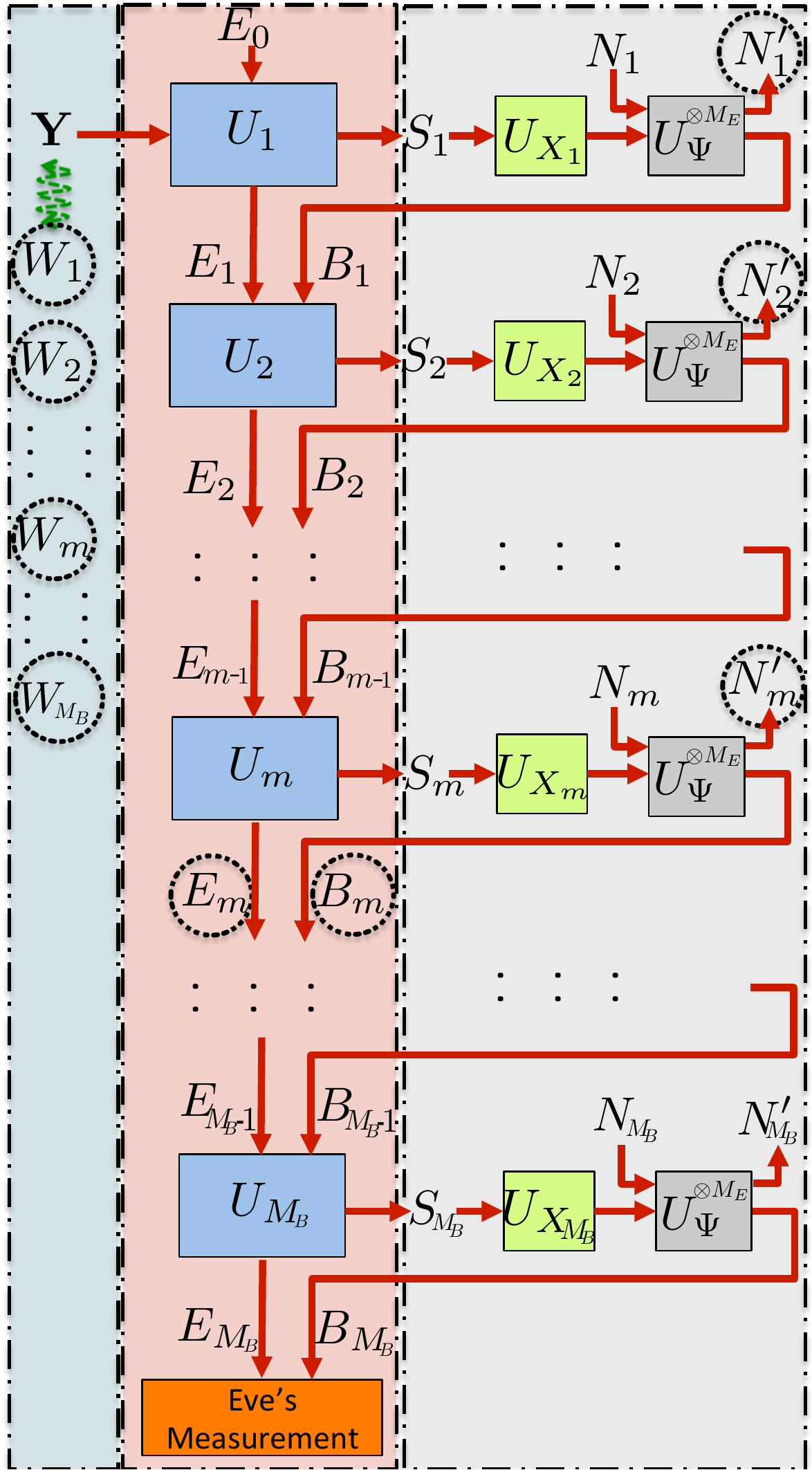}
\caption{Schematic of the most general coherent attack on an $M_B$-symbol block.  Dotted circles enclose the modes present after the $m$th attack round.}
\label{most_general}
\end{figure}
\subsection{Bounding the block-wise coherent attack}
\label{block_bound}
Reference~\cite{Zhuang_2016_cl} developed a way to bound Eve's information gain for a TW-QKD protocol, but that reference focused on a rigorous formulation for noisy entanglement-assisted classical capacity, and did not present in full detail a security proof for TW-QKD protocols.  Here we present a detailed proof that Ref.~\cite{Zhuang_2016_cl}'s capacity formula provides an upper bound on the the most general block-wise coherent attack's information gain.  Our proof applies to \emph{all} TW-QKD protocols in which Bob performs the key map, not just to the Gaussian special case.  As shown in Fig.~\ref{most_general}, Alice transmits the $M_B$-symbol block ${\bm Y}\equiv Y_1\cdots Y_{M_B}$, where the $\{Y_m\}$ each have $M_E$ modes and we are using the Fig.~\ref{scheme_QKD} notation with a subscript to identify the symbol's place within the $M_B$-symbol block.
Alice has access to the purifications ${\bm W} \equiv W_1\cdots W_{M_B}$, i.e., each $(Y_mW_m)$ pair is in the tensor product of $M_E$ TMSV states.  

We shall allow Eve to perform the most general attack on this block---shown schematically in Fig.~\ref{most_general}---by supposing that Alice sends all of ${\bm Y}$ simultaneously on the forward channel, and that Eve performs the following $M_B$-round interactive process with Bob.  In the $m$th round, Eve sends an $M_E$-mode signal $S_m$~\cite{footnote3} to Bob who responds by transmitting his $M_E$-mode encoded symbol $B_m$~\cite{footnote4}.  We assume that Eve captures $B_m$ in its entirety, because doing so will aid our security analysis and full capture affords Eve more information than she would get from partial capture.  In addition, we will give Eve an ideal quantum memory, so that she can postpone her quantum measurement until after the $M_B$th interaction round.  Furthermore, we grant Eve the power to create an arbitrarily entangled multi-mode ancilla $E_0$ for use in her first round, and the ability to perform an arbitrary multi-mode unitary, $U_m$, in the $m$th round.  Her first round's unitary acts on ${\bm Y}$ and produces the outputs $S_1$, which Eve sends to Bob, and $E_1$, which is an ancilla that Eve retains for use in the second round.  In the next $M_B-2$ rounds, Eve's unitary acts on $(E_mB_m)$ and produces the output $E_{m+1}$.  Eve makes her quantum measurement on $(E_{M_B}B_{M_B})$, the outputs from her $M_B$th interaction round.  

Bob's operations in Fig.~\ref{most_general} are the following.  Upon receipt of $S_m$ from Eve, he encodes a randomly-chosen classical symbol $X_m$ by means of the unitary $U_{X_m}$ which he then passes through the channel $\Psi^{\otimes M_E}$, whose Stinespring-dilation unitary $U_\Psi^{\otimes M_E}$ has ancilla input $N_m$ and produces the return $B_m$ and the ancilla output $N_m^\prime$.  (Note that Fig.~\ref{most_general} has omitted Bob's use of a small portion of each $S_m$ for security testing.)

We assume that Eve can neither access the ${\bm W}$ that are in Alice's lab nor the ${\bm N} \equiv N_1\cdots N_{M_B}$ and the ${\bm N}' \equiv N_1^\prime\cdots N_{M_B}^\prime$ that are in Bob's lab.
Thus her Holevo-information gain satisfies
\begin{align}
\chi_E^{(M)}&\equiv I(E_{M_B}B_{M_B}: X_{M_B}\cdots X_1)
\\
&=\sum_{m=1}^{M_B} I(E_{M_B}B_{M_B}: X_m \mid X_{m-1}\cdots X_1)
\\
&\le \sum_{m=1}^{M_B} I(E_mB_m: X_m \mid X_{m-1}\cdots X_1),
\label{chiE_all}
\end{align}
where the superscript $M = M_BM_E$ denotes the total number of modes from which Eve has gained information.
The first equality is because the Holevo information obtainable from a quantum system $A$, in state $\rho_A^x$ with probability density function $p_X(x)$, about a classical register $X$ can be written as the Shannon information $I(A:X)$ between $A$ and $X$ for the classical-quantum state $\rho_{AX}\equiv\int\!{\rm d}x\, p_X(x)\rho_A^x \otimes  \ket{x}_X\bra{x}$~\cite{nielsen2010quantum}. The second equality is due to the chain rule for Shannon information, with $I(A:C\mid B)$ being the conditional Shannon information between $A$ and $C$ given $B$~\cite{fawzi2015quantum}. The inequality follows from the data processing inequality~\cite{nielsen2010quantum}, because key distribution after the $m$th round is a quantum operation that generates $E_{M_B}B_{M_B}$ from $E_mB_m$ with assistance from ancilla that are independent of $X_m$, and hence do not increase the Shannon information. 

Now let us focus on the system after $m$th round, which consists of $E_m B_m W_1\cdots W_{M_B} N_1^\prime \cdots N_m^\prime $.  These modes, which are contained in Fig.~\ref{most_general}'s dotted circles, are in a joint pure state with Eve only having access to $E_m B_m$.   Nevertheless, it is convenient to \emph{increase} Eve's information gain by pretending that she can access $B_m$ and $\tilde{E}_m\equiv E_m W_1\cdots W_{m-1}W_{m+1} W_{M_B} N_1^\prime \cdots N_{m-1}^\prime$, i.e., we have that 
\ba
&&I(E_mB_m: X_m \mid X_{m-1}\cdots X_1)
\nonumber
\\
&&\le I(\tilde{E}_mB_m: X_m \mid X_{m-1}\cdots X_1).
\label{MB_block_chi}
\ea
The term on the right in (\ref{MB_block_chi}) is bounded above by the noisy entanglement-assisted capacity formula from Ref.~\cite{Zhuang_2016_cl}, with $\tilde{E}_m$ as the ancilla generated by Eve and $W_m \tilde{E}_m S_m$ in a pure state. Thus we have $\chi_E^{(M)}$ from~(\ref{chiE_all}) has an upper bound given by $M_B$ times a multi-letter capacity formula over $M_E$ modes, where we emphasize that this result applies to \emph{all} TW-QKD protocols in which Bob performs the key map.  

\begin{figure}[bp]
\includegraphics[width=0.45\textwidth]{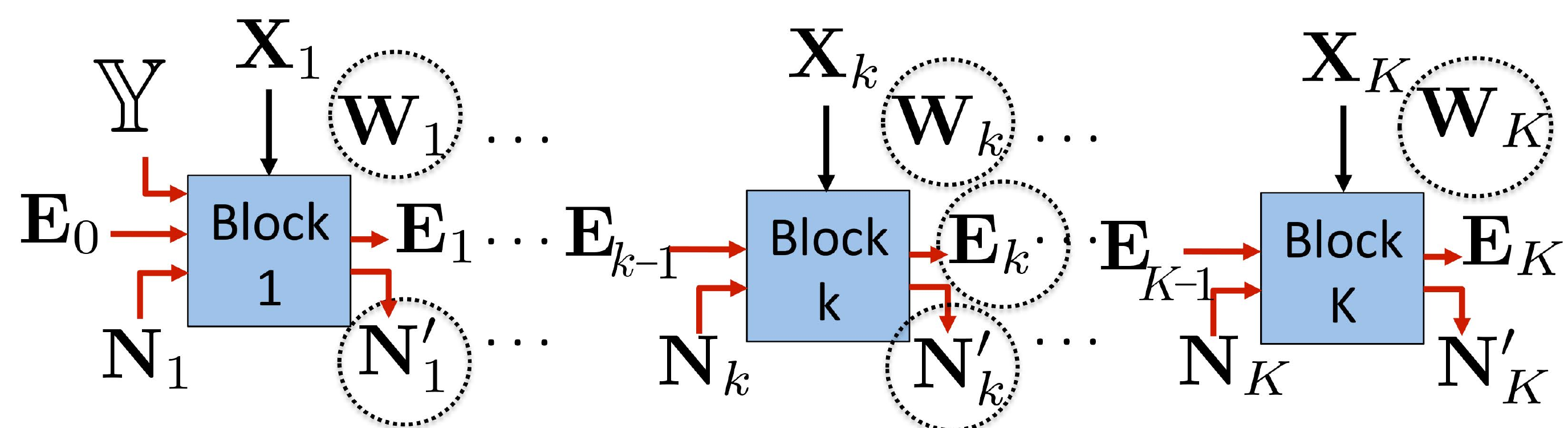}
\caption{Schematic of the most general coherent attack on $K$ blocks of $M_B$ symbols. The dotted circles enclose the modes present after $k$th block. Note that all the $\{\bf W_k\}$ have been present from the start, despite their being assigned to different $k$ values in this figure.}
\label{most_general_block}
\end{figure}
\subsection{Optimality of the block-wise coherent attack}
\label{block_reduce}
Despite the de~Finetti theorem sufficing to reduce the coherent attack to the block-wise coherent attack, here we provide an analysis that the block-wise coherent attack is the optimum coherent attack in the asymptotic regime. Note that the analysis is not entirely rigorous yet, however, there is good hope that with some future generalization of quantum asymptotic-equipartition property (QAEP)~\cite{tomamichel2009fully}, the analysis will be fully rigorous. This approach is desirable, not only because it is more elegant, but also since we expect it will lead to tighter finite-key bounds. Consider $K$ blocks of $M_B$ symbols that are indexed by $1\le k\le K$. The schematic for Eve's coherent attack on these $K$ blocks is similar to the single-block attack from Fig.~\ref{most_general}, except that now all of Alice's signal modes, $\mathbb{Y}\equiv{\bm Y}^{(1)}\cdots {\bm Y}^{(K)}$, are supplied to Eve simultaneously.  To proceed expeditiously, we introduce some new notation. In the $k$th block, we let $\bm E_k\equiv E_{M_B}^{(k)}B_{M_B}^{(k)}$, $\bm N_k\equiv N_1^{(k)}\cdots N_{M_B}^{(k)}$, $\bm N_k^\prime\equiv N_1^{\prime(k)}\cdots N_{M_B}^{\prime(k)}$, and $\bm W_k\equiv W_1^{(k)}\cdots W_{M_B}^{(k)}$, where, except for the superscript denoting the block index, the right-hand sides of each definition have the same meanings as in Fig.~\ref{most_general}.   In a similar manner, we use $\bm X_k\equiv X_1^{(k)}\cdots X_{M_B}^{(k)}$ to denote Bob's random classical messages for the $k$th block.  

With the preceding notation, Fig.~\ref{most_general_block} shows the schematic for Eve's $K$-block coherent attack, in which the $k$th block can be considered a unitary from input $\bm E_{k-1} \bm N_k$ to output $\bm E_k \bm N_k^\prime$, conditioned on the classical messages $\bm X_k$. For a $K$-block QKD session, the $\epsilon_{\rm EC}+\epsilon_{\rm PA}+\epsilon$-secure SKE, in bits/symbol, is given by~\cite{renner2008security}
%\begin{eqnarray}
%{\rm SKE}(K,M_B) &=& [H_{\rm min}^\epsilon(\bm X_1\cdots \bm X_K \mid \bm E_K)
%\nonumber
%\\[.05in]
%&&\,\,-{\rm leak}_{\rm IR}+\log(\epsilon_{\rm PA}^2)]/KM_B,
%\label{Finite_rate}
%\end{eqnarray}
\begin{align}
&{\rm SKE}(K,M_B) 
\nonumber
\\
&= \left[H_{\rm min}^\epsilon\left(\bm X_1\cdots \bm X_K \mid \bm E_K\right)
-{\rm leak}_{\rm IR}+\log\left(\epsilon_{\rm PA}^2\right)\right]/KM_B,
\label{Finite_rate}
\end{align}
where ${\rm leak}_{\rm IR}$ is the information leaked to Eve in the information reconciliation protocol with $\epsilon_{\rm EC}$-secure error correction and $\epsilon_{\rm PA}$-secure privacy amplification, and $H_{\rm min}^\epsilon(A \mid B)$ is the smooth min-entropy of $A$ conditioned on $B$. Note that ${\rm leak}_{\rm IR}$ can be determined by Alice and Bob. 
In the asymptotic ($K\rightarrow \infty$) regime the last term in brackets vanishes, so we only need to lower bound $H_{\rm min}^\epsilon(\bm X_1\cdots \bm X_K \mid \bm E_K)$ for Alice and Bob to have a lower bound on their SKE.  The arguments that follow parallel the single-block case. 

First, we use the chain rule for smooth min-entropy~\cite{vitanov2013chain} repeatedly to obtain
\begin{align}
&H_{\rm min}^\epsilon(\bm X_1\cdots \bm X_K \mid \bm E_K)
\ge H_{\rm min}^{z}(\bm X_K \mid \bm E_K)+
\nonumber
\\
&
\sum_{k=1}^{K-1} H_{\rm min}^{z}(\bm X_k \mid \bm E_K \bm X_{k+1}\cdots \bm X_K)-(K-1)f(z),
\label{H_1}
\end{align}
where $f(z)\sim \log(1/z)$ and $z=\epsilon/(3K-2)$. Because $\bm E_K \bm X_{k+1}\cdots \bm X_K$ can be obtained from $\bm E_k$ by a quantum operation, the data-processing inequality for smooth min-entropy~\cite{renner2008security} gives us
\be
H_{\rm min}^{z}(\bm X_k\mid \bm E_K \bm X_{k+1}\cdots \bm X_K) \ge H_{\rm min}^{z}(\bm X_k \mid \bm E_k).
\label{H_2}
\ee
Next, after the $k$th block, we decrease Eve's smooth min-entropy by granting her access to everything other than $\bm W_k$ and $\bm N_k^\prime$, i.e., Eve's system is enlarged to $\tilde{\bm E}_k\equiv \bm E_k\bm W_1\cdots \bm W_{k-1}\bm W_{k+1}\cdots \bm W_{K} \bm N_1^\prime \cdots \bm N_{k-1}^\prime$. Another use of the data-processing inequality then leads to 
\be
H_{\rm min}^{z}(\bm X_k \mid \bm E_k )\ge H_{\rm min}^{z}(\bm X_k \mid \tilde{\bm E}_k),
\label{H_3}
\ee
where the right-hand side corresponds to the case in which Eve has a pure state $k$th block's outset. Combining Eqs.~(\ref{Finite_rate})--(\ref{H_3}), we get
\ba
{\rm SKE}(K,M_B)\ge \left[
\sum_{k=1}^K H_{\rm min}^{z}(\bm X_k \mid \tilde{\bm E}_k)
\right.
\nonumber
\\[.05in]
-(K-1)f(z)-{\rm leak}_{\rm IR}+\log(\epsilon_{\rm PA}^2)]/KM_B.
\label{Finite_rate2}
\ea
Because $f(z)\sim \log(3K/\epsilon)$, we can let $K$ and $M_B$ increase while maintaining $K\gg M_B\gg \log(K)\gg 1$, and obtain, asymptotically, 
\ba
\lefteqn{{\rm SKE}(K,M_B) \ge} \nonumber \\[.05in]
&& \frac{1}{KM_B}\left[\sum_{k=1}^{K} H_{\rm min}^{\epsilon/3K}(\bm X_k \mid \tilde{\bm E}_k)-{\rm leak}_{\rm IR}\right].
\label{Finite_rate3}
\ea

The preceding lower bound is achieved when Eve performs independent operations on each $M_B$-symbol block. If Alice and Bob's security testing leads to identical constraints on each block, then the asymptotic-regime lower bound is achieved by Eve's performing a block-wise coherent attack. In Eve's absence, QKD protocols operating with $M_B \gg 1$ give security-testing results that are nearly identical for all sufficiently-large blocks. When Eve's activities create substantial block-to-block variations in Alice and Bob's security-testing results, they abort the protocol.

However, in order to be fully rigorous, one still need to show that (\ref{Finite_rate3}) is lower bounded by the bound in (\ref{DW_rate}). However, the QAEP in ref.~\cite{tomamichel2009fully} does not apply directly, since there is no independent and identically distributed structure in (\ref{Finite_rate3}). To close the last step, one would require generalization of QAEP, which is a future direction to pursue.
Conditioned on the QAEP generalization, as was the case for the bound in~(\ref{MB_block_chi}), the bound in~(\ref{Finite_rate3}) and its implications apply to \emph{all} TW-QKD protocols. Part of our analysis is similar to the idea of entropy accumulation~\cite{dupuis2016entropy}, which has been successfully applied in device-independent QKD protocols~\cite{arnon2018practical,arnon2016simple}. However, owing to the structure of TW-QKD, in which Eve can interactively alter the quantum states being sent between Alice and Bob, the framework of entropy accumulation does not apply directly to our problem.

\subsection{Constraints and single-letterization}
Here we assume a Gaussian TW-QKD protocol and return to (\ref{chiE_all}) and (\ref{MB_block_chi}), in which $\chi_E^{(M)}$ in (\ref{chiE_all}) is bounded above by $M_B$ times the multi-letter capacity formula from Ref.~\cite{Zhuang_2016_cl} across $M_E$ modes. For simplicity, however, we will use the multi-letter capacity formula from Ref.~\cite{Zhuang_2016_cl} across $M=M_BM_E$ modes, which still establishes an upper bound on $\chi_E^{(M)}$. 
Going forward, we will use $\bm S\equiv S_1S_2\cdots S_M$, $\bm B\equiv B_1B_2\cdots B_{M}$, and $\bm W\equiv W_1W_2\cdots W_M$ to denote the modes involved. 
For Gaussian protocols, $U_X$ is covariant with $\Psi$, thus Eve's information gain obeys the following multi-letter bound~\cite{Zhuang_2016_cl,devetak2005distillation},
\ba
\chi_E^{(M)}&\le&\max_{\rho_{{\bm S}{\bm W}}}
F(\rho_{{\bm S}{\bm W}}),
\label{capacity_Eve}
\\[.05in]
F(\rho_{{\bm S}{\bm W}})&\equiv&
S(\rho_{\bm B})-E_{{(\Psi^{\otimes M})}^c\otimes \mathcal{I}}(\rho_{{\bm S}{\bm W}}).
\label{capacity_covariant}
\ea
In this bound: $S(\cdot)$ is the von Neumann entropy; $\phi^c$ denotes the complementary channel to the $\phi$ channel; $\mathcal{I}$ is the identity channel on $\bm W$; and $E_\phi(\cdot)$, the entropy gain of the completely-positive trace-preserving map $\phi$ applied to a system in state $\rho$,  is defined to be 
$E_\phi(\rho)\equiv S[\phi(\rho)]-S(\rho)$. 
The maximization in~(\ref{capacity_Eve}) is over attacks that are constrained by the intrusion parameters that Alice and Bob derive from their security testing on the state $\rho_{{\bm S}{\bm W}}$.   We shall assume, in proceeding, that Bob independently encodes each mode ($M_E=1$), so that $\rho_{B_m}=\int\!{\rm d}x\, p_X(x) \Psi(U_x \rho_{S_m}U_x^\dagger)$; when Bob uses $M_E > 1$ encoding, (\ref{capacity_Eve}) is still an upper bound on $\chi_E^{(M)}$. 

To facilitate evaluating (\ref{capacity_Eve}), and thus the asymptotic-regime SKE from (\ref{DW_rate}), the constraints that Alice and Bob derive from their security testing should satisfy two requirements. 
\begin{enumerate}[label=\text{(R\arabic*)},wide, labelwidth=!,labelindent=0pt]
\item \label{requirement1}
The constraints lead to a single-letter upper bounds on Eve's information gain. 
\item \label{requirement2}
The constraints can be measured precisely in the asymptotic regime from Alice and Bob's performing an LOCC procedure. 
\end{enumerate}
Requirement~\ref{requirement1} ensures that evaluating the upper bound on Eve's information gain from the constraints is tractable, and requirement~\ref{requirement2} ensures that the constraints can be obtained with arbitrarily high precision from security testing over a sufficiently long QKD session.  

Because $\rho_{{\bm S}{\bm W}}$ is infinite dimensional, we use Gaussian extremality~\cite{wolf2006extremality,garcia2006unconditional}---which states that when the covariance matrix of the input state is fixed, continuous sub-additive (super-additive) function, which is invariant under local passive symplectic transforms, has its maximum (minimum) achieved by Gaussian states---to satisfy requirement~\ref{requirement1} and restrict the maximization in (\ref{capacity_Eve}) to Gaussian states.
Toward this end, Ref.~\cite{Zhuang_2016_cl} established the following two sub-additivity inequalities (Theorems~2 and~3 in Ref.~\cite{Zhuang_2016_cl}'s supplemental material),
\ba
&F(\rho_{{\bm S}{\bm W}})
\le  
\sum_{m=1}^M  F(\rho_{S_m W_m}),
\label{ineq1}
\\[.05in]
&F(\rho_{{\bm S}{\bm W}})
\le  
\sum_{m=1}^M  F(\rho_{S_m{\bm W}}).
\label{ineq2}
\ea
Here $F(\rho_{S_mW_m})\equiv
S(\rho_{B_m})-E_{\Psi^c\otimes \mathcal{I}}(\rho_{S_mW_m})$ and $F(\rho_{S_m{\bm W}})\equiv S(\rho_{B_m})-E_{\Psi^c\otimes \mathcal{I}}(\rho_{S_m{\bm W}})$ are generalizations of Eq.~(\ref{capacity_covariant}).
Because $E_\phi$ is convex~\cite{Holevo2011}, we have that $F(\rho_{{\bm S}{\bm W}})$ is concave in quantum states. The subadditivity inequalities~(\ref{ineq1}) and~(\ref{ineq2}) then ensure that the maximum in~(\ref{capacity_Eve}) is achieved by Gaussian inputs, $\rho_{{\bm S}{\bm W}}$~\cite{Zhuang_2016_cl}, that satisfy covariance-matrix constraints. Thus we will only consider constraints from security testing that restrict covariance matrices.

In Sec.~\ref{Sec_separate}, we revisit the covariance-matrix constraints considered in Ref.~\cite{Zhuang_2016_cl}, and give its explicit form for Gaussian protocols.  Although these constraints meet requirement~\ref{requirement1}, they fail to satisfy requirement~\ref{requirement2}, making them unsuitable for our goal of establishing a TW-QKD security framework.   In Sec.~\ref{Sec_mode_wise}, we introduce constraints---in the form of sums of pair-wise terms---and show that they meet requirements~\ref{requirement1} \emph{and}~\ref{requirement2}. Similarly, in Sec.~\ref{section_general_form_sum}, we generalize to sum constraints that are invariant under signal-mode permutations, and show that they too obey requirements~\ref{requirement1} and~\ref{requirement2}.  Under a collective attack, the precision with which the intrusion parameters from Secs.~\ref{Sec_mode_wise} and~\ref{section_general_form_sum} can be estimated improves as the QKD session's duration increases, becoming perfect in the asymptotic limit. Moreover, standard CV-QKD covariance-estimation techniques can be applied for that purpose.  It is an important and open problem, however, to find means for reliable estimation of these intrusion parameters when Eve performs a coherent attack.  A procedure that would suffice in that regard is one that affords a robust measurement of $\overline{\boldsymbol \Lambda} \equiv \sum_{m=1}^M{\boldsymbol \Lambda}_{S_mW_m}/M$, where ${\boldsymbol \Lambda}_{S_mW_m}$ is the Wigner covariance matrix of $(S_mW_m)$.  

There are two reasons why the single-letter bounds on Eve's coherent-attack information gain that result from using~(\ref{ineq1}) or~(\ref{ineq2}) in~(\ref{capacity_Eve}) may not be tight:  (1) they assume that Eve collects all the light that Bob sends to Alice; and (2) they assume single-mode encoding.  The first reason does not apply to long-distance QKD, because security analysis presumes Eve collects all the light lost in propagation from Bob to Alice, and that loss is 90\% for a 50-km-long low-loss (0.2\,dB/km) fiber and 99\% for a 100-km-long fiber.  Even for short-haul links the first reason does not apply to FL-QKD, because that protocol employs a high-gain optical amplifier in Bob's terminal.  In contrast, FL-QKD employs multi-mode encoding with $M_E \gg 1$~\cite{Quntao_2015,zhuang2017large,zhang2016floodlight,zhang2017floodlight}, whereas the protocol from Ref.~\cite{Pirandola_2008} uses single-mode encoding, so the latter is immune to the second reason although it is prone to the first.  

\subsubsection{Separate pair-wise constraints}
\label{Sec_separate}

Reference~\cite{Zhuang_2016_cl} imposed pair-wise constraints on the reduced density operators, $\{\rho_{S_m W_m} : 1\le n\le M\}$, to reduce~(\ref{capacity_Eve}) to a single-letter formula via~(\ref{ineq1}). To be specific, suppose that, when Eve mounts her attack, Alice and Bob's security-testing measurements allows them to determine the average photon numbers of all the $\{S_m\}$ modes, 
\be
\braket{\hat{a}^\dagger_{S_m}\hat{a}_{S_m}}=\kappa_S^{(m)}N_S,\mbox{ for $1\le m \le M$},
\label{Q1}
\ee
and the total cross-correlation strengths for all $(S_mW_m)$ pairs, 
\be
|\braket{\hat{a}_{S_m}\hat{a}_{W_m}}|^2+|\braket{\hat{a}^\dagger_{S_m}\hat{a}_{W_m}}|^2=\kappa_f^{(m)} C_S^2, \mbox{ for $1\le m \le M$}.
\label{Q2}
\ee
The intrusion parameters $\{\kappa_S^{(m)}\}$ quantify the average photon numbers of Bob's $\{S_m\}$ modes relative to $N_S$, the average photon number of the $\{Y_m\}$ modes that Alice transmitted, while the intrusion parameters $\{\kappa_f^{(m)}\}$ quantify the total cross-correlation strengths of the $\{(S_mW_m)\}$ pairs relative to those of the $\{(Y_mW_m)\}$.  In order for these parameters to be physically valid, we require that $\kappa_S^{(m)} \ge 0$ and  
$0\le \kappa_f^{(m)}\le \min[\kappa_S^{(m)},(1+2\kappa_S^{(m)}N_S)/(1+2N_S)]$, as shown in Appendix~\ref{AppA}.

Using (\ref{ineq1}), the information-gain bound in (\ref{capacity_Eve}) reduces to a single-letter form~\cite{Zhuang_2016_cl}
\be
\chi_E^{(M)} \le \sum_{m=1}^M \chi_E\!\left(\kappa_S^{(m)},\kappa_f^{(m)}\right),
\label{single_sum}
\ee
where
\be
\chi_E\!\left(\kappa_S^{(m)},\kappa_f^{(m)}\right)\equiv\max_{\rho_{S_mW_m}} F(\rho_{S_mW_m}),
\label{single_term}
\ee
with the maximization being constrained by the intrusion parameters from Eqs.~(\ref{Q1}) and~(\ref{Q2}). 
The following theorem guarantees that Eq.~(\ref{single_term}) is easily evaluated. 
\begin{theorem}
For Gaussian TW-QKD protocols, with intrusion parameters given by Eqs.~(\ref{Q1}) and~(\ref{Q2}), the maximization in Eq.~(\ref{single_term}) results in
\begin{align}
\chi_E\!\left(\kappa_S^{(m)},\kappa_f^{(m)}\right)&=g(N_{B_m})-
E_{\Psi^c\otimes \mathcal{I}}^\star\!\left(\kappa_S^{(m)},\kappa_f^{(m)}\right),
\label{final_opt_main}
\end{align}
where $g(N_T) = (N_T+1)\log_2(n_T+1)-N_T\log_2(N_T)$ is the von Neumann entropy of a thermal state with average photon number $N_T$, $N_{B_m} = \langle \hat{a}_{B_m}^\dagger\hat{a}_{B_m}\rangle$ from Eq.~(\ref{Nb}) for the $m$th mode, and $E_{\Psi^c\otimes \mathcal{I}}^\star\!\left(\kappa_S,\kappa_f\right)$ is a minimized entropy gain that can be evaluated as a three-parameter minimization of a closed-form analytic function.  

When the intrusion parameters satisfy $\kappa_f^{(m)}\simeq \kappa_S^{(m)}\le1$, the maximum is achieved by the beam-splitter light injection attack that was shown in Ref.~\cite{Quntao_2015} to realize Eve's optimum frequency-domain collective attack.

\label{theorem_opt_attack}
\end{theorem}
The proof of Theorem~1 is similar to the frequency-domain collective attack proof from Ref.~\cite{Quntao_2015}; see Appendix~\ref{AppA} for the details.  
We emphasize that strong numerical evidence (see Appendix~\ref{AppA}) suggests that the beam-splitter light injection attack is the optimum attack when $\kappa_f^{(m)}\le (1+\kappa_SN_S)/(1+N_S)$ for both the quantum-limited amplifier channel, $\mathcal{A}_{G_B}^{0}$, and its complementary channel, $\tilde{\mathcal{A}}_{G_B}^{0}$, and when $\kappa_S\le1$ for the pure-loss channel, $\mathcal{L}_{\eta_B}^{0}$.

Unfortunately, when Eve mounts a coherent attack each $\rho_{S_mW_m}$ may be different, which implies that Alice and Bob only get a single instance of that state from which it is impossible to get reliable estimates of $\kappa_S^{(m)}$ and $\kappa_f^{(m)}$.  Thus the separate pair-wise constraints fail to satisfy requirement~\ref{requirement2}. 

\subsubsection{Pair-wise sum constraint}
\label{Sec_mode_wise}
As a first approach to remedying the separate pair-wise constraint's robustness deficiency, let us consider the  pair-wise sum constraints, 
\be
\sum_{m=1}^M
\braket{\hat{a}^\dagger_{S_m}\hat{a}_{S_m}}=M\overline{\kappa}_S N_S,
\label{Q1s}
\ee
and
\be
\sum_{m=1}^M 
\left[{|\braket{\hat{a}_{S_m}\hat{a}_{W_m}}|^2+|\braket{\hat{a}^\dagger_{S_m}\hat{a}_{W_m}}|^2}\right]=M\overline{\kappa}_f C_S^2,
\label{Q2s}
\ee
which are so named because they are the sums of quantities involving only a single mode pair.

In turns out, as we now show, that the pair-wise sum constraints' intrusion parameters $\overline{\kappa}_S$ and $\overline{\kappa}_f$ allow the information-gain bound in~(\ref{capacity_Eve}) to be reduced to the single-letter formula
\be
\chi_E^{(M)} \le 
M \chi_E(\overline{\kappa}_S,\overline{\kappa}_f),
\label{capacity_Eve_pair_wise}
\ee
where $\chi_E(\overline{\kappa}_S,\overline{\kappa}_f)$ is obtained from Eq.~(\ref{single_term}) with $\kappa_S^{(m)} = \overline{\kappa}_S$ and $\kappa_f^{(m)} = \overline{\kappa}_f$.    To demonstrate that this is so, let us first
suppose  $\chi_E\!\left(\kappa_S^{(m)},\kappa_f^{(m)}\right)$ is a concave function, in which case we have that 
$\chi_E^{(M)} \le \sum_{n=1}^M \chi_E\!\left(\kappa_S^{(n)},\kappa_f^{(n)}\right)\le 
M \chi_E(\overline{\kappa}_S,\overline{\kappa}_f).$  The second inequality becomes an equality when the mode pairs are independent and identically distributed. Moreover, given the average Wigner covariance matrix, $\overline{\boldsymbol \Lambda}$, we can obtain $\overline{\kappa}_S$ because $\sum_{m=1}^M\langle \hat{a}_{S_m}^\dagger\hat{a}_{S_m}\rangle/M = \overline{\kappa}_SN_S$ is one of $\overline{\boldsymbol \Lambda}$'s diagonal elements.  We can also get a lower bound on $\overline{\kappa}_f$ from $\overline{\boldsymbol \Lambda}$, because $\overline{\boldsymbol \Lambda}$'s off-diagonal elements obey $|\sum_{m=1}^M\langle \hat{a}_{S_m}\hat{a}_{W_m}\rangle|^2/M + |\sum_{m=1}^M\langle \hat{a}^\dagger_{S_m}\hat{a}_{W_m}\rangle|^2/M \le \overline{\kappa}_fC_S^2$.  
Then, because $\chi_E(\overline{\kappa}_S,\overline{\kappa}_f)$ increases with decreasing $\overline{\kappa}_f$ for fixed $\kappa_S$, we can use the intrusion parameters derived from $\overline{\boldsymbol \Lambda}$ in  (\ref{capacity_Eve_pair_wise}) to bound Eve's information gain.

To complete our demonstration that the pair-wise sum constraints provide an upper bound on Eve's information gain, we must verify that $\chi_E\!\left(\kappa_S^{(m)},\kappa_f^{(m)}\right)$ from Eq.~(\ref{final_opt_main}) is concave.    
The von Neumann entropy is concave, so all that needs to be shown is that $E_{\Psi^c\otimes \mathcal{I}}^\star\!\left(\kappa_S^{(m)},\kappa_f^{(m)}\right)$ is convex.  That term is the constrained minimum of an entropy gain whose lengthy closed-form expression makes is difficult to prove the desired convexity analytically.  Our numerical work in Appendix~\ref{AppA}, however, indicates that Eq.~(\ref{final_opt_main}) is indeed a concave function of $(\kappa_S^{(m)},\kappa_f^{(m)})$ for $\mathcal{A}_{G_B}^{0}$, $\mathcal{L}_{\eta_B}^{0}$, and $\tilde{\mathcal{A}}_{G_B}^{0}$ (see Fig.~\ref{concavity_fig}).  In practice, Alice and Bob's protocol will operate in the vicinity of some nominal set of intrusion parameters, so our numerical evidence should suffice for justifying the use of pair-wise sum constraints.  

\subsubsection{Permutation-invariant sum constraints}
\label{section_general_form_sum}

The pair-wise sum constraints' cross-correlation intrusion parameter, $\overline{\kappa}_f$ from Eq.~(\ref{Q2s}), 
may be difficult to measure when, as in FL-QKD, Bob uses multi-mode encoding with $M_E \gg 1$.  In this section, therefore, we will replace Eq.~(\ref{Q2s})'s cross-correlation constraint with the permutation-invariant constraint, 
\be
{\sum_{m,n=1}^M \left[|\braket{\hat{a}_{S_m}\hat{a}_{W_n}}|^2+|\braket{\hat{a}^\dagger_{S_m}\hat{a}_{W_n}}|^2\right]}=\overline{K}_f {MC_S^2},
\label{Cgeneral_form1}
\ee
which constrains the total cross correlation between the $\{\hat{a}_{S_m}\}$ modes and all permutations of the $\{\hat{a}_{W_m}\}$ modes.  Its measurement may be easier than that for the pair-wise sum constraint when $M_E \neq 1$.  
Because $\overline{K}_f\ge \overline{\kappa}_f$, a lower bound for $\overline{K}_f$ can also be obtained from the average covariance matrix $\overline{\boldsymbol \Lambda}$.  

We now show that with $\overline{\kappa}_S$ and $\overline{K}_f$ from Eqs.~(\ref{Q1s}) and~(\ref{Cgeneral_form1}) we get the information-gain upper bound
\be
\chi_E^{(M)}\le M  \chi_E(\overline{\kappa}_S,\overline{K}_f).
\label{capacity_Eve_invariant}
\ee
To do so, we reduce the permutation-invariant sum constraints to the separate mode-pair constraints, Eqs.~(\ref{Q1}) and~(\ref{Q2}), as follows.
We start by using~(\ref{ineq2}) in (\ref{capacity_Eve}) so that the maximization to be done is of $\sum_{m=1}^M  F(\rho_{S_m\bm W})$.  Next, we introduce an intermediate intrusion parameter, $K_f^{(m)}$, defined by
\be
{\sum_{n=1}^M\left[|\braket{\hat{a}_{S_m}\hat{a}_{W_n}}|^2+|\braket{\hat{a}^\dagger_{S_m}\hat{a}_{W_n}}|^2\right]}=K_f^{(m)} C_S^2,
\label{Q2_general}
\ee
so that Eq.~(\ref{Cgeneral_form1}) can be rewritten as $\sum_{m=1}^M K_f^{(m)}/M=\overline{K}_f.$
An upper bound on the maximum of $\sum_{m=1}^M  F(\rho_{S_m\bm W})$ can thus be obtained in two steps. First, for fixed $\kappa_S^{(m)},K_f^{(m)}$, obtain the maximum of $ F(\rho_{S_m\bm W})$ over $\rho_{S_m\bm W}$. Then maximize over the set $\{\kappa_S^{(m)},K_f^{(m)} : 1\le m\le M\}$. The first maximization is accomplished by the following theorem.
\begin{theorem}
For a Gaussian TW-QKD protocol with references modes ${\bm W}$ and a signal mode $S_m$, we have that  
\begin{eqnarray}
\lefteqn{\chi'_E\!\left(\kappa_S^{(m)},K_f^{(m)}\right) \equiv \max_{\rho_{S_mW_1W_2}}F(\rho_{S_m\bm W})} \\[.05in]
&=& \max_{\rho_{S_mW_1W_2}} 
[S(\rho_B)-E_{\Psi^c\otimes \mathcal{I}}(\rho_{S_mW_1W_2})].
\label{goal_proof_2_main}
\end{eqnarray}
under the Eq.~(\ref{Q1}) constraint and 
\be
{|\braket{\hat{a}_{S_m}\hat{a}_{W_1}}|^2+|\braket{\hat{a}_{S_m}^\dagger\hat{a}_{W_1}}|^2+|\braket{\hat{a}_{S_m}\hat{a}_{W_2}}|^2}=K_f^{(m)} C_S^2,
\label{non_zeros_main}
\ee 
where the maximization in Eq.~(\ref{goal_proof_2_main}) can be accomplished by a four-parameter maximization of a closed-form analytic function.
\label{theorem_opt_attack_general}
\end{theorem}
The proof's basic idea is to manipulate the ${\bm W}$ modes with properly chosen beam splitters; see Appendix~\ref{AppB} for the details.  Unfortunately, the four-parameter maximization is analytically cumbersome, because of the lengthy nature of the closed-form expression involved. 
Consequently we again resort to numerics. As shown in Appendix~\ref{AppB}, we find that for the $\mathcal{A}_{G_B}^{0}$ and $\tilde{\mathcal{A}}_{G_B}^{0}$ channels with various $G_B$ values, as well as for the $\mathcal{L}_{\eta_B}^{0}$ channel with various $\eta_B$ values, the maximum is achieved, for various $N_S$ values, when $|\braket{\hat{a}_{S_m}^\dagger\hat{a}_{W_1}}|^2=0$.  At this point, suitable beam splitting of the $\bm W$ modes can make $\braket{\hat{a}_{S_m}\hat{a}_{W_2}}=0$.  This collapses the Eq.~(\ref{non_zeros_main}) constraint to the single-mode pair constraint in Eq.~(\ref{Q2}), giving us
$\chi'_E\!\left(\kappa_S^{(m)},K_f^{(m)}\right)=\chi_E\!\left(\kappa_S^{(m)},K_f^{(m)}\right)$. Combined with concavity arguments, we obtain the information-gain bound in~(\ref{capacity_Eve_invariant}).

\begin{figure}
\includegraphics[width=0.48\textwidth]{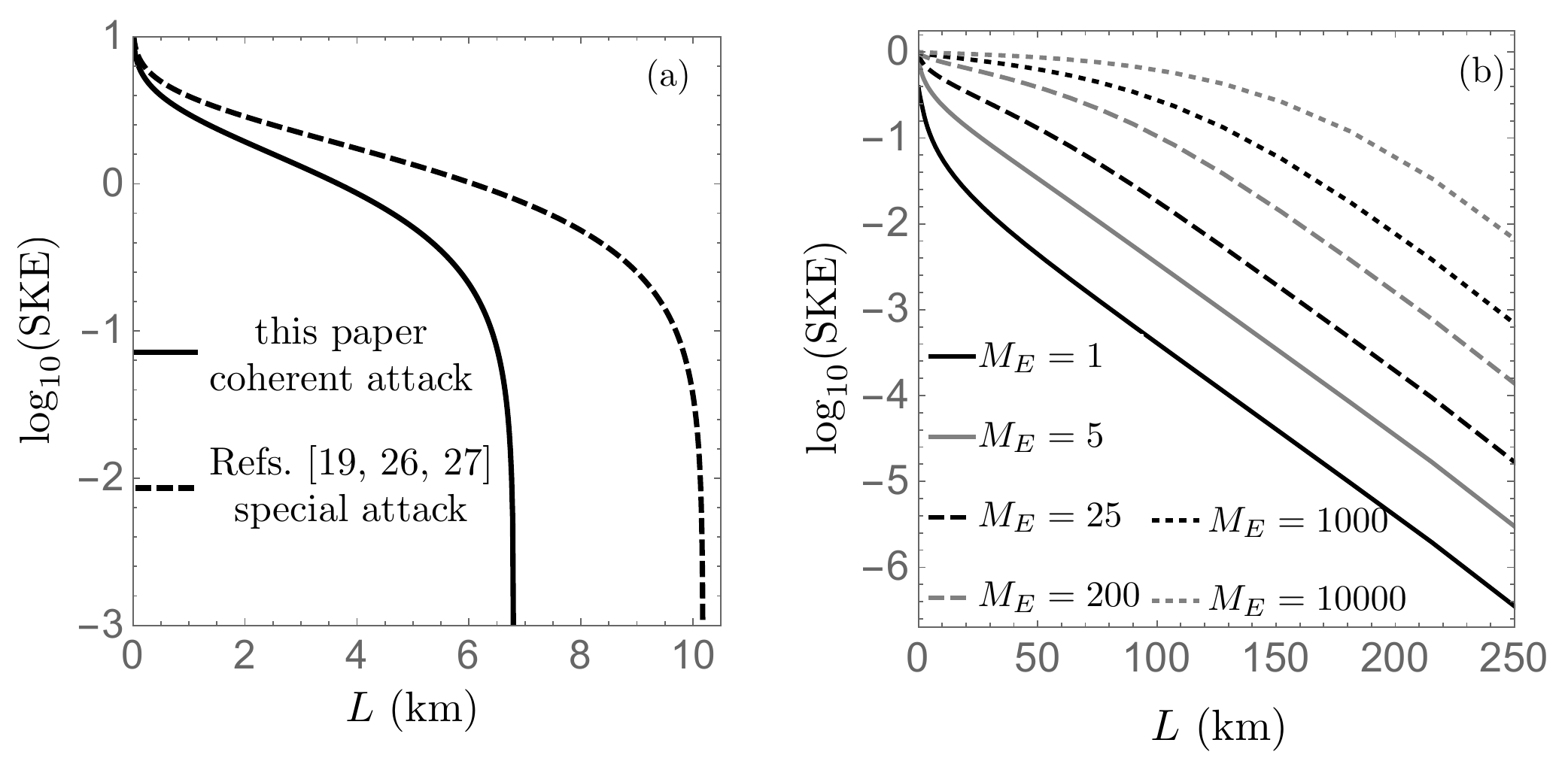}
\caption{Secret-key efficiencies in bits/symbol versus one-way path length (channel transmissivity $\kappa_S = 10^{-0.02L}$) for attacks that give $\overline{\kappa}_S = \kappa_S$ and $\overline{\kappa}_f = \kappa_S$ and post-processing that gives perfect reconciliation efficiency.  The solid curves are coherent-attack SKE lower bounds obtained from this paper's framework.  (a) Results for the TMSV protocol with $N_S \gg 1$ and $E_X \gg 1$; the dashed curve is the SKE lower bound from Refs.~\cite{Pirandola_2008,Ottaviani_2015,Ottaviani_2016}.  (b)  Results for the FL-QKD protocol with $G_B=10^6$, $N_S$ chosen at each $L$ to maximize the SKE, and various $M_E$ values.}
\label{rate}
\end{figure}

\section{Secret-key efficiencies}
\label{SKR_various}
In this section, we evaluate the asymptotic SKEs, given by Eq.~(\ref{DW_rate}) under Eve's coherent attack, for two Gaussian TW-QKD protocols:  the TMSV protocol from Refs.~\cite{Pirandola_2008,Weedbrook_2014,Ottaviani_2015,Ottaviani_2016}, and FL-QKD~\cite{Quntao_2015,zhuang2017large,zhang2016floodlight,zhang2017floodlight}.
These protocols' SKRs can be obtained, if desired, by multiplying their SKEs by Bob's encoding rate, e.g., $R=10$\,Gbaud for state-of-the art equipment. 
We assume that asymptotic-regime operation permits the intrusion parameters $\overline{\kappa}_S, \overline{\kappa}_f$ or $\overline{\kappa}_S, \overline{K}_f$ to be measured perfectly, and using those parameters we can bound Eve's information gain per mode using $\chi_E(\overline{\kappa}_S,\overline{\kappa}_f)$ or $\chi_E(\overline{\kappa}_S,\overline{K}_f)$ in Eq.~(\ref{final_opt_main}). With that result in hand we can get a lower bound on Alice and Bob's SKE once we have evaluated their Shannon information in bits/symbol.   For that evaluation we need to specify Bob's encoding operation and Alice's measurement on the light each receives in the protocol under consideration.  In what follows we will do so assuming that the Alice-to-Bob and Bob-to-Alice channels, in the absence of Eve, are optical-fiber links with 0.2\,dB/km loss, so that $\kappa_S = 10^{-0.02L}$, where $L$ is the one-way distance in km between Alice and Bob.  We shall neglect the additional losses associated with Alice and Bob's security testing.

\subsection{TMSV protocol with random displacement}
\label{protocol1}
In the TMSV protocol with random displacement~\cite{Pirandola_2008,Ottaviani_2015,Ottaviani_2016}, Alice has access to the full TMSV state and Bob performs single-mode encoding using zero-mean, circulo-complex, Gaussian-distributed displacements that add average photon number $E_X$ to each mode he receives.  Bob does not employ an additional operation after his encoding, thus $\Psi$ is the noiseless identity channel that is equivalent to $\mathcal{A}_{1}^{0}$, and hence $N_B=\kappa_SN_S+E_X$ from Eq.~(\ref{Nb}) in the absence of Eve, or when her attack does not alter Bob's average received photon number.  Alice uses a dual-homodyne receiver to measure both quadratures of the light she receives.  Given the intrusion parameters $\overline{\kappa}_S$ and $\overline{\kappa}_f$ from Eqs.~(\ref{Q1s}) and~(\ref{Q2s}), our framework provides asymptotic security for this protocol against coherent attacks.

To illustrate the SKEs predicted by our framework for the TMSV protocol, we consider an attack---like Eve's passive attack in which she only interacts with the light lost in propagation between Alice and Bob and between Bob and Alice---that preserves Alice and Bob's covariance matrix, so that $\overline{\kappa}_S = \kappa_S$ and $\overline{\kappa}_f = \kappa_S$.  Under this attack, Alice and Bob's Shannon information is~\cite{Ottaviani_2015}
$I_{AB}=\log_2 (\kappa_SE_X+\kappa_S^2 N_S+1),$
and their resulting SKE is 
\be 
{\rm SKE} = \max[ I_{AB}-\chi_E(\kappa_S, \kappa_S),0],
\label{DW_rate_single}
\ee 
where we have assumed perfect reconciliation efficiency, $\beta =1$.  

Figure~\ref{rate}(a) plots our coherent-attack SKE and the special-attack SKE from Refs.~\cite{Pirandola_2008,Ottaviani_2015,Ottaviani_2016} versus the one-way path length, $L$, where we have taken $N_S\gg 1$ and $E_X\gg 1$, which makes this results independent of the exact values of those system parameters.  This figure shows our SKE prediction to be much lower than the previous result.  This gap is primarily due to our giving Eve access to all the light on the Bob-to-Alice channel which, for the short distances over which the TMSV protocol operates, is overly conservative, viz., $\kappa_S=0.63$ at $L=10$\,km.  Indeed, for Eve's passive attack the SKE from Refs.~\cite{Pirandola_2008,Ottaviani_2015,Ottaviani_2016} \emph{is} the TMSV protocol's true performance. But our framework provides an SKE lower bound for an arbitrary coherent attack---which can result in $\overline{\kappa}_S \neq \kappa_S$ and $\overline{\kappa}_f \neq \kappa_S$---whereas the SKE result from Refs.~\cite{Pirandola_2008,Ottaviani_2015,Ottaviani_2016} does not.   

We expect our SKE lower bound to be much tighter for more robust protocols' long-distance operation, wherein $\kappa_S \ll 1$. Moreover, we might tighten our SKE bound for short-distance protocols by adding security testing on the Bob-to-Alice channel.  For example, Bob might merge signal light from his own TMSV source with his $B$ mode in his transmission to Alice while retaining that source's idler light for them to use in an LOCC procedure that will provide intrusion parameters quantifying Eve's intrusion on the Bob-to-Alice channel.  

\subsection{FL-QKD protocol}
\label{protocol2}
FL-QKD~\cite{Quntao_2015,zhang2016floodlight,zhang2017floodlight} is a two-way continuous-variable QKD protocol.  Alice transmits unmodulated light to Bob.  Bob binary-phase-shift encodes ($\theta_X = 0$ or $\pi$\,rad, $d_X = 0$) that light with random bits~\cite{footnote5}, and sends the encoded light back to Alice, who homodyne detects what she receives.   FL-QKD introduces two major innovations:  (1) Alice transmits broadband amplified spontaneous emission (ASE) light to Bob at low brightness ($< 1$ photon/mode) while retaining a high-brightness ($\gg 1$ photon/mode) version for use as her homodyne receiver's local oscillator.  (2) Bob sends his encoded version of the light he received from Alice through a high-gain optical amplifier ($\Psi = \mathcal{A}^0_{G_B}$ with $G_B \gg1$) before transmission back to Alice.  

The preceding innovations completely defeat passive eavesdropping and enable FL-QKD to achieve Gbit/s SKRs against such an attack for the following reasons:  (1) Alice's low-brightness transmission, after Bob's encoding operation, gets buried in the ASE noise of his high-gain amplifier.  This noise makes it impossible to retrieve Bob's bit string without a high-brightness replica of the light Alice sent to Bob.  Alice has such a reference, but the no-cloning theorem precludes Eve's generating one from Alice's low-brightness transmission. (2) Bob's encoding rate ($R \sim 10$\,Gbit/s) is much lower than the bandwidth ($W_B \sim 2$\,THz) of Alice's ASE transmission.  The resulting high value of the bit-time $\times$ optical-bandwidth product ($W_B/R \sim 200$) enables Alice to send many photons per bit time to Bob, thus mitigating the Alice-to-Bob channel's loss in the same manner as in classical optical communication.  (3) Bob's high-gain amplifier can completely overcome the Bob-to-Alice channel's loss.  Consequently, FL-QKD is a two-way protocol whose effective propagation loss is that of one-way transmission.  

Were \emph{passive} eavesdropping the only threat faced by FL-QKD, its protocol security would be completely assured.  Like other two-way protocols, however, FL-QKD is vulnerable to an \emph{active} eavesdropping attack, in which Eve shines her own light into Bob's terminal---while saving her own reference beam---and then determines his bit string by using her reference to detect his encoding of her illumination from light she culls from the Bob-to-Alice channel.  FL-QKD has been shown---both theoretically~\cite{Quntao_2015} and experimentally~\cite{zhang2016floodlight,zhang2017floodlight} to defeat active eavesdropping by channel monitoring that uses a very low brightness photon-pair source at Alice's terminal, together with photon-counting measurements at both Alice and Bob's terminals, to bound the amount light Eve has injected into Bob.  In fact, this monitoring, whose photon-pair source is a spontaneous parametric downconverter that produces multi-mode TMSV states, provides security against the optimum frequency-domain collective attack~\cite{Quntao_2015}.  A great virtue of the present paper is that its framework can be applied to ensure FL-QKD's security against a coherent attack, as we now show.

Because $\bm Y$, Alice's transmission to Bob, merges her low-brightness ASE light with the signal beam from her SPDC source, Alice only has access to \emph{part} of $\bm W$, the purification of that transmission.  That part, ${\bm W}'$, is her SPDC source's idler beam that she retains for use in security testing.  Nevertheless, that retained light suffices for our asymptotic-regime security framework, because $\langle \hat{a}_{S_m}\hat{a}_{W'_n}\rangle = \sqrt{\tau}\,\langle \hat{a}_{S_m}\hat{a}_{W_n}\rangle$ and $\langle \hat{a}^\dagger_{S_m}\hat{a}_{W'_n}\rangle = \sqrt{\tau}\,\langle \hat{a}^\dagger_{S_m}\hat{a}_{W_n}\rangle$ for all $m,n$, where $\tau$ is the fraction of Alice's transmission to Bob that is due to her SPDC source.   
Thus, with Alice and Bob determining their Shannon information from error-probability measurements, and assuming that they can obtain the intrusion parameters $\overline{\kappa}_S$ and $\overline{K}_f$, they have what they need to set a lower bound on the asymptotic-regime, coherent-attack SKE.  

Our final task will be to illustrate the behavior of that bound when Eve's attack does not impact Alice and Bob's covariance matrix, so that $\overline{\kappa}_S = \kappa_S$ and $\overline{K}_f = \kappa_S$, and their reconciliation efficiency is perfect, $\beta =1$~\cite{footnote6}.  In this case they have an assured SKE that satisfies
\be 
{\rm SKE} = \max[ I_{AB}-M_E\,\chi_E(\kappa_S,\kappa_S),0].
\label{DW_rate_M}
\ee
This SKE is plotted versus one-way path length in Fig.~\ref{rate}(b) for various $M_E$ values with $I_{AB}$ obtained from Alice and Bob's theoretical error probability~\cite{Quntao_2015}, $G_B = 10^6$, and source brightness, $N_S$, chosen at each $L$ to maximize SKE.  For $M_E = 200$ and $R = 10\,$Gbit/s, Fig.~\ref{rate}(b) predicts an ${\rm SKR} = R\,{\rm SKE}$ in excess of 2\,Gbit/s at $L = 50\,$km, as found for those parameter values in our previous frequency-domain collective attack security analysis~\cite{Quntao_2015} with the equivalent of $\overline{\kappa}_S = \kappa_S$, $\overline{K}_f = 0.99\,\kappa_S$, and $\beta = 0.94$.    

Figure~\ref{rate}(b) also underscores the value of multi-mode encoding in achieving high SKEs, and hence high bits/s SKRs for a given symbol rate $R$.  All QKD protocols have bits/mode SKRs bounded above by the PLOB bound~\cite{pirandola2017fundamental}, $-\log_2(1-\kappa_S)$\,bits/mode.  Figure~\ref{rate}(b)'s single-mode encoding ($M_E = 1$) curve is well \emph{below} that bound, but its $M_E \gg 1$ curves report bits/symbol rates that are well \emph{above} $-\log_2(1-\kappa_S)$.  This is why, for the same symbol rate $R$, FL-QKD can realize much higher bits/s SKRs than the predominant decoy-state BB84 protocol, because the latter employs single-mode encoding and its state-of-the-art implementation~\cite{lucamarini2013efficient} has bits/mode performance on par with Fig.~\ref{rate}(b)'s $M_E = 1$ curve at 50\,km one-way path length.

\section{Summary and discussion}
\label{discussion}
In this paper we have taken significant steps toward an asymptotic-regime, coherent-attack security proof for TW-QKD protocols.  First, we showed that the noisy entanglement-assisted channel capacity formula~\cite{Zhuang_2016_cl} provides an upper bound on Eve's information gain from her most general coherent attack.  Then, we exhibited covariance-matrix constraints that can provide efficiently calculable bounds on her information gain for Gaussian TW-QKD protocols, and showed that the resulting upper bound can be achieved by a collective attack.  Finally, we applied our results to two such protocols, the TMSV protocol~\cite{Pirandola_2008,Weedbrook_2014,Ottaviani_2015,Ottaviani_2016} and FL-QKD~\cite{Quntao_2015,zhuang2017large,zhang2016floodlight,zhang2017floodlight}.   The latter example is especially important, because FL-QKD offers the potential for Gbit/s SKRs over metropolitan-area distances without the need for any new technology but its current security analysis only assures protection against a frequency-domain collective attack~\cite{Quntao_2015}.  As a result, developing LOCC security tests that will permit Alice and Bob to obtain the intrusion parameters employed in our framework is open problem of great significance.  These parameters can, in principle, be obtained from standard homodyne measurements when Eve's attack is collective, rather than coherent, and Alice and Bob's QKD protocol uses single-mode encoding, but a measurement approach that works for coherent attacks on multi-mode encoding is needed.  One possibility may be 
to use the reliable state tomography technique~\cite{christandl2012reliable}. Note that the pairwise-sum constraint in Eq.~(\ref{Q2s}) is \emph{not} invariant to the basis chosen by Alice and Bob for their modes~\cite{footnote7}.  But Alice and Bob only need to measure this constraint in a particular basis, to bound Eve's information gain, as long as the mode transformations within Alice and Bob's equipment are described by the channels from Sec.~\ref{intro_TWQKD}.  The permutation-invariant sum constraint in Eq.~(\ref{Cgeneral_form1}), on the other hand, \emph{is} invariant to the choice of basis, because it can be written in terms correlations of the continuous-time field operators, $\hat{E}_S(t)$ and $\hat{E}_W(t)$, for Bob's received signal and Alice's purification~\cite{footnote7}.  Even if one or both of the preceding constraints can be measured, the general composite-security, finite-key analysis for Eve's coherent attack will still need to be worked out for FL-QKD and other Gaussian TW-QKD protocols.

Finally, we must emphasize that our security-proof framework's goal is to establish the \emph{protocol} security of TW-QKD.  It does not address such protocols' \emph{implementation} security, i.e., side-channel attacks that exploit device characteristics---including deviations from their normal operating regimes---to compromise key exchange.  That said, QKD still offers implementation security that is independent of future technological advances:  any attack must be executed with the technology that is available at the time of the key exchange.

\begin{acknowledgements}
QZ, ZZ, and JHS acknowledge support from Air Force Office of Scientific Research Grant No.~FA9550-14-1-0052.  JHS also acknowledges support from Office of Naval Research Contract No.~N00014-16-C-2069, and  QZ also acknowledges support from the Claude E. Shannon Research Assistantship.  NL acknowledges support by NSERC under the Discovery Program. The Institute for Quantum Computing is supported by  the Government of Canada and the Province of Ontario. QZ thanks the Perimeter Institute for its hospitality, Rotem Arnon-Friedman for discussions of entropy accumulation, and Felix Leditzky for discussions.
\end{acknowledgements}

\appendix

\section{Proof of Theorem~\ref{theorem_opt_attack} \label{AppA}}
Because Theorem~\ref{theorem_opt_attack} deals with a single mode-pair, we shall omit mode-index superscripts and subscripts and employ the notation from Fig.~\ref{scheme_QKD} throughout what follows.  Thus our 
objective is to show that $F(\rho_{SW})\equiv
S(\rho_{B})-E_{\Psi^c\otimes \mathcal{I}}(\rho_{ SW})$, when maximized over states $\rho_{SW}$ satisfying
\ba
&\braket{\hat{a}^\dagger_{S}\hat{a}_{S}}=\kappa_SN_S,
\label{Q1_proof}
\\
&{|\braket{\hat{a}_{S}\hat{a}_{W}}|^2+|\braket{\hat{a}^\dagger_{S}\hat{a}_{W}}|^2}=\kappa_f C_S^2,
\label{Q2_proof}
\ea
obeys
\be 
\chi_E(\kappa_S,\kappa_f) = g(N_B) - E_{\Psi^c\otimes \mathcal{I}}^\star(\kappa_S,\kappa_f).
\ee
Here, $g(N_B)$ is the von Neumann entropy of a thermal state with average photon number $N_B$ where $N_B$ is given by Eq.(\ref{Nb}), and $E_{\Psi^c\otimes \mathcal{I}}^\star(\kappa_S,\kappa_f)$ is the entropy gain minimized over the preceding constraints.

Before proceeding with the details, let us outline the structure of the proof.  Equations~(\ref{Q1_proof}) and~(\ref{Q2_proof}) are functions of $\rho_{SW}$'s covariance matrix ${\boldsymbol \Lambda}_{SW}$. The subadditivity of $F(\rho_{SW})$ therefore implies that the constrained maximum is achieved by a Gaussian-state $\rho_{SW}$~\cite{Zhuang_2016_cl}. Thus we need only consider an eavesdropper's using a Gaussian unitary, namely a $(K+1)$-mode Bogoliubov transformation~\cite{Weedbrook_2012} parameterized by a set of variables.  Owing to Eq.~(\ref{Nb}), $S(\rho_B)$ is bounded above by $g(N_B)$. To complete the proof we need a non-trivial lower bound on $E_{\Psi^c\otimes \mathcal{I}}(\rho_{SW})=S(\rho_{N'W})-S(\rho_{SW})$, where $N'$ is the environment mode after Bob's Gaussian channel $\Psi$.  Moreover, to obtain that bound we only need the covariance matrices ${\boldsymbol \Lambda}_{SW}$ and ${\boldsymbol \Lambda}_{N'W}$ of $\rho_{SW}$ and $\rho_{N^\prime W}$, which can be found from Alice's ${\boldsymbol \Lambda}_{YW}$ and the parameters of Eve's Bogoliubov transformation. Then $E_{\Psi^c\otimes \mathcal{I}}(\rho_{SW})$ is obtained by a symplectic diagonalization that turns out to depend on only three parameters of the Bogoliubov transformation, given the constraints Eqs.~(\ref{Q1_proof}) and~(\ref{Q2_proof}). Minimizing over these three parameters yields $E_{\Psi^c\otimes \mathcal{I}}^\star(\kappa_S,\kappa_f)$. However, $E_{\Psi^c\otimes \mathcal{I}}(\rho_{SW})$'s closed-form expression is rather complicated, which prevents analytical minimization, hence we will rely on numerical minimization. That said, we will use series expansion in the vicinity of $\kappa_f = \kappa_S$ to show that the beam-splitter light injection attack~\cite{Quntao_2015} is always a local minimum in that region, wherein Alice and Bob's security testing has severely limited Eve's intrusion.\\  
 
\begin{proof} 
Let Eve's Gaussian unitary---her $K+1$-mode Bogoliubov transformation~\cite{Weedbrook_2012}---be, 
\begin{equation}
\hat{a}_{S} = u_0\hat{a}_{Y} + v_0^*\hat{a}_{Y}^\dagger + \sum_{k=1}^K (u_k\hat{e}_{V}^{(k)} + v_k^*\hat{e}_{V}^{(k)\dagger} )+ \alpha.
\label{Bogoliubov}
\end{equation}
where $\hat{a}_Y$ is the photon annihilation operator of Alice's $Y$ mode, and $\{\,\hat{e}^{(k)} : 1\le k \le K\,\}$ are the photon annihilation operators of Eve's ancilla modes, all of which are in their vacuum states.  
We require Eq.~(\ref{Bogoliubov}) to yield a proper free-field commutator bracket for $\hat{a}_{S}$, thus the complex-valued coefficients $\{\, u_k,v_k : 0 \le k \le K\,\}$ must satisfy
\begin{equation}
|u_0|^2+{\bf u}^\dagger{\bf u}-|v_0|^2-{\bf v}^\dagger{\bf v} = 1,
\label{commutatorpreservation}
\end{equation}
where ${\bf u}^\dagger = \left[\begin{array}{cccc} u_1^* & u_2^* & \cdots & u_K^*\end{array}\right]$, with $^\dagger$ denoting conjugate transpose, and a similar definition for ${\bf v}^\dagger$.  Equations~(\ref{Q1_proof}) and~(\ref{Q2_proof}) impose their own restrictions on $\{u_k,v_k,\alpha\}$:  
\ba
|\alpha|^2+|v_0|^2+{\bf v}^\dagger{\bf v}&=&(\kappa_S-\kappa_f)N_S.
\label{Q1p}
\\
|u_0|^2+|v_0|^2&=&\kappa_f,
\label{Q2p}
\ea

We will maximize $S(\rho_B)$ and minimize $E_{\Psi^c\otimes \mathcal{I}}(\rho_{SW})$ separately, and show that they can be achieved simultaneously. 

\subsection{Maximizing $S(\rho_B)$}
\label{entropy_first}
Here we show that Eve's Gaussian unitary with $\alpha = 0$ achieves the constrained maximization of $S(\rho_B)$.  Because $\rho_B$ is a displaced thermal state, we know that 
\be
\max_{\rho_{SW}} 
S(\rho_B)=g(N_B-|\alpha|^2),
\label{goal_proof_SB_result}
\ee
where $N_B$ is given by Eq.~(\ref{Nb}) and the $\{u_k,v_k,\alpha\}$ satisfy Eqs.~(\ref{commutatorpreservation})--(\ref{Q2p}),   
which implies that $\alpha=0$ maximizes $S(\rho_B)$ under the given constraints.  Furthermore, because the entropy-gain term is independent of the displacement $\alpha$, we have that $\chi_E(\kappa_S,\kappa_f)$ is achieved by $\alpha=0$.  

\subsection{Minimizing $E_{\Psi^c\otimes \mathcal{I}}(\rho_{SW})$}
\label{entropy_gain_term}
Here we perform the constrained minimization,
\be
\min_{\rho_{SW}} 
E_{\Psi^c\otimes \mathcal{I}}(\rho_{SW})
\equiv
\min_{\rho_{SW}}  S(\rho_{N'W})-S(\rho_{SW}),
\label{goal_proof_En}
\ee
where $\rho_{N'W}=\Psi^c\otimes \mathcal{I}(\rho_{SW})$ is the joint state of the environment and the purification after Bob's channel $\Psi$. Because $\rho_{N'W}$ and $\rho_{SW}$ are Gaussian, their entropies are given, in terms of their covariance matrices' symplectic eigenvalues---$\{\nu_{\pm}\}$ for ${\boldsymbol \Lambda}_{N'W}$ and $\{\mu_{\pm}\}$ for ${\boldsymbol \Lambda}_{SW}$)---which leads to 
\begin{align}
E_{\Psi^c\otimes \mathcal{I}}(\rho_{N'W})&=g[(4\nu_+-1)/2] +g[(4\nu_--1)/2] 
\nonumber\\[.05in]
&-g[(4\mu_+-1)/2]  - g[(4\mu_--1)/2].
\label{munu_exp}
\end{align}
where the symplectic eigenvalues must satisfy Eqs.~(\ref{commutatorpreservation})--(\ref{Q2p}) with $\alpha = 0$.  

Maximizing Eq.~(\ref{munu_exp}) over the $\{u_k,v_k\}$ is more readily accomplished by rewriting Eqs.~(\ref{commutatorpreservation})--(\ref{Q2p}) in terms of $\{\gamma,\delta,\theta_v,\theta_{uv}\}$ chosen such that
\begin{subequations}
\begin{align}
u_0 &= \sqrt{\kappa_f}\,\sin(\gamma), 
\label{u0}\\[.05in]
v_0 &= \sqrt{\kappa_f}\,\cos(\gamma)e^{i\theta_v}, \label{v0}\\[.05in]
{\bf u}^\dagger{\bf u} &= (\kappa_S-\kappa_f)N_S + 1-\kappa_f+\kappa_f\cos^2(\gamma),\label{unorm}\\[.05in]
{\bf v}^\dagger{\bf v} &= (\kappa_S-\kappa_f)N_S -\kappa_f\cos^2(\gamma),\label{vnorm}\\[.05in]
{\bf v}^\dagger{\bf u} &= \sqrt{({\bf v}^\dagger{\bf v})({\bf u}^\dagger{\bf u})}\,\cos(\delta)e^{i\theta_{uv}}.
\label{vu}
\end{align}
\label{para_all}
\end{subequations}
In these expressions:  $\gamma \in [0,\pi/2]$ satisfies
\be
1-\frac{1}{\kappa_f}-\left(\frac{\kappa_S}{\kappa_f}-1\right)\!N_S\le \cos^2(\gamma) \le \left(\frac{\kappa_S}{\kappa_f}-1\right)\!N_S;
\ee
$\delta\in [0,\pi/2]$; and $u_0$ has been taken to be non-negative, without loss of generality, because global phase is irrelevant.  

The foregoing reformulation makes it easy to show that \emph{all} states $\rho_{SW}$ must have $\kappa_S$ and $\kappa_f$, defined by Eqs.~(\ref{Q1}) and (\ref{Q2}), that satisfy
\be
0\le \kappa_f\le \min[\kappa_S,(1+2\kappa_SN_S)/(1+2N_S)].
\label{kf_exist}
\ee
Specifically: $\kappa_f \ge 0$ follows from its definition in Eq.~(\ref{Q2}); $\kappa_f \le \kappa_S$ follows from $(\kappa_S - \kappa_f)N_S = |v_0|^2 + {\bf v}^\dagger{\bf v} \ge 0$; $\kappa_f \le  (1+2\kappa_SN_S)/(1+2N_S)$ follows from $2(\kappa_S-\kappa_f)N_S + 1-\kappa_f = {\bf u}^\dagger{\bf u} + {\bf v}^\dagger{\bf v} \ge 0$; and the generality of the result is because the $\kappa_f$ limits apply to the covariance matrix of an arbitrary, not just a Gaussian, $\rho_{SW}$.   
 
\subsubsection{Covariance matrix of $\hat{a}_S$ and  $\hat{a}_W$}
Equations~(\ref{cov_TMSV}) and~(\ref{Bogoliubov}) enable us to show that the covariance of $\hat{a}_S$ and $\hat{a}_W$ is given by 
\begin{equation}
{\boldsymbol \Lambda}_{SW} = \frac{1}{4}\!\left[\begin{array}{ccc} {\bf A}_S& & {\bf C}_{SW} \\[.05in]
{\bf C}_{SW} & & {\bf A}_{W}\end{array}\right],
\end{equation}
where:  
\begin{equation}
{\bf A}_{S} =  2\!\left[\begin{array}{cc} A_S + {\rm Re}(w)&  
{\rm Im}(w)\\[.05in]
{\rm Im}(w)&  A_S - {\rm Re}(w) 
\end{array}\right],
\end{equation}
with $A_S = 1/2 + \kappa_S N_S$ and $w ={\bf v}^\dagger{\bf u} + (2N_S+1)v_0^*u_0$; 
\begin{equation}
{\bf C}_{SW} = 2C_S\left[\begin{array}{ccc} u_0 + {\rm Re}(v_0) & & {\rm Im}(v_0) \\[.05in]
-{\rm Im}(v_0) & & -u_0 +{\rm Re}(v_0)\end{array}\right],
\end{equation}
with $C_S = \sqrt{N_S(N_S+1)}$; and ${\bf A}_W = (2N_S +1){\bf I}_2$.

\subsubsection{Covariance matrix of $\hat{a}_{N}^\prime$ and $\hat{a}_W$}
Because Bob's encoding, $U_X$, is covariant with his channel, $\Psi$,  we can omit $U_X$ in calculating ${\boldsymbol \Lambda}_{N'W}$ for Bob's three channels, i.e., his quantum-limited amplifier channel ($\Psi = \mathcal{A}_{G_B}^{0}$), his pure-loss channel ($\Psi = \mathcal{L}_{\eta_B}^{0}$), and his contravariant quantum-limited amplifier channel ($\Psi = \tilde{\mathcal{A}}_{G_B}^{0}$). 
\label{Cov_NS}
\begin{enumerate}
\item
Quantum-limited amplifier channel, with
$\hat{a}_N^\prime = \sqrt{G_B-1}\,\hat{a}_S^\dagger + \sqrt{G_B}\,\hat{a}_N$
and $G_B\ge 1$. Here we have that 
\begin{equation}
{\boldsymbol \Lambda}_{N'W} = \frac{1}{4}\!\left[\begin{array}{ccc} {\bf A}_{N'} & & {\bf C}_{N'W} \\[.05in]
{\bf C}_{N'W} & & {\bf A}_W\end{array}\right],
\end{equation}
where: 
\begin{equation}
{\bf A}_{N'} =  2\left[\begin{array}{ccc}A' + {\rm Re}(x)  & & -{\rm Im}(x) \\[.05in]
-{\rm Im}(x)& & A' -{\rm Re}(x) \end{array}\right],
\end{equation}
with $A' = 1/2 + G_BN_B + (G_B-1)(\kappa_SN_S + 1)$ and $x = (G_B-1)w$; and 
\begin{eqnarray}
{\bf C}_{N'W} &=& 2\sqrt{G_B-1}\,C_S \nonumber \\[.05in]
&\times& \left[\begin{array}{ccc} u_0 + {\rm Re}(v_0) & &  {\rm Im}(v_0) \\[.05in]
{\rm Im}(v_0) & & u_0 - {\rm Re}(v_0)\end{array}\right]. 
\end{eqnarray} 

\item
Pure-loss channel with $\hat{a}_N^\prime= \sqrt{1-\eta_B}\,\hat{a}_S - \sqrt{\eta_B}\,\hat{a}_N$ and 
$0\le \eta_B\le 1$. Here we find that 
\begin{equation}
{\boldsymbol \Lambda}_{N'W} = \frac{1}{4}\!\left[\begin{array}{ccc} {\bf A}_N' & & {\bf C}_{N'W} \\[.05in]
{\bf C}_{N'W} & & {\bf A}_W\end{array}\right],
\end{equation}
where:  
\begin{equation}
{\bf A}_{N'} =  2\left[\begin{array}{ccc}A' + {\rm Re}(x)  & & {\rm Im}(x) \\[.05in]
{\rm Im}(x)& & A' -{\rm Re}(x) \end{array}\right],\\[.05in]
\end{equation}
with $A' = 1/2 +(1-\eta_B)N_S + \eta_BN_B$ and $x= (1-\eta_B)w$;
and
\begin{equation}
{\bf C}_{N'W} = \sqrt{1-\eta_B} {\bf C}_{SW}. 
\end{equation} 

\item
Contravariant quantum-limited amplifier channel with $\hat{a}_N^\prime =\sqrt{G_B}\,\hat{a}_S+\sqrt{G_B-1}\,\hat{a}_N^\dagger$ and $G_B \ge 1$. Now we get
\begin{equation}
{\boldsymbol \Lambda}_{N'W} = \frac{1}{4}\!\left[\begin{array}{ccc} {\bf A}_{N'} & & {\bf C}_{N'W} \\[.05in]
{\bf C}_{N'W} & & {\bf A}_W\end{array}\right],
\end{equation}
where: 
\begin{equation}
{\bf A}_{N'} =  2\left[\begin{array}{ccc}A' + {\rm Re}(x)  & & {\rm Im}(x) \\[.05in]
{\rm Im}(x)& & A' -{\rm Re}(x) \end{array}\right],
\end{equation}
with $A' = 1/2+G_B\kappa_SN_S+(G_B-1) (N_B+1)$ and $x = G_Bw$;
and
\begin{equation}
{\bf C}_{N'W} = \sqrt{G_B}\,{\bf C}_{SW}. 
\end{equation} 

\end{enumerate}

\begin{figure*}
\includegraphics[width=0.98\textwidth]{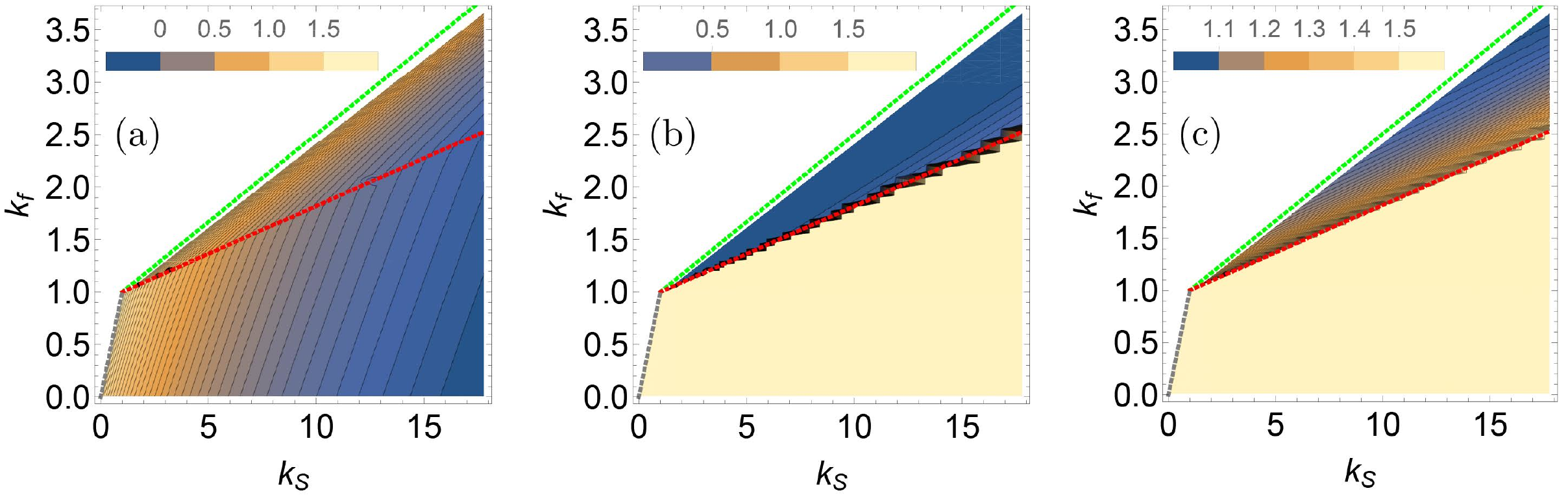}\
\caption{Numerically-obtained entropy-gain minimization results for the quantum-limited amplifier channel with $N_S=0.1$ and $G_B=1.5$.  (a) Minimum entropy gain, $E_{\Psi^c\otimes \mathcal{I}}^\star(\kappa_S,\kappa_f)$.  (b) Optimum $\delta$ value.  (c) Optimum $\gamma$ value.  In (b) and (c) the green line, $\kappa_f = (1+2\kappa_SN_S)/(1+2N_S)$, and the gray line, $\kappa_f = \kappa_S$, mark the (\ref{kf_exist}) upper limit on possible $\kappa_f$ values, and the red line, $\kappa_f = (1+\kappa_SN_S)/(1+N_S)$, is the $\kappa_f$ value below which the local minimum at $\gamma = \delta = \pi/2$ is also the global minimum.}
\label{minimization_Amp}
\end{figure*}

\begin{figure*}
\includegraphics[width=0.98\textwidth]{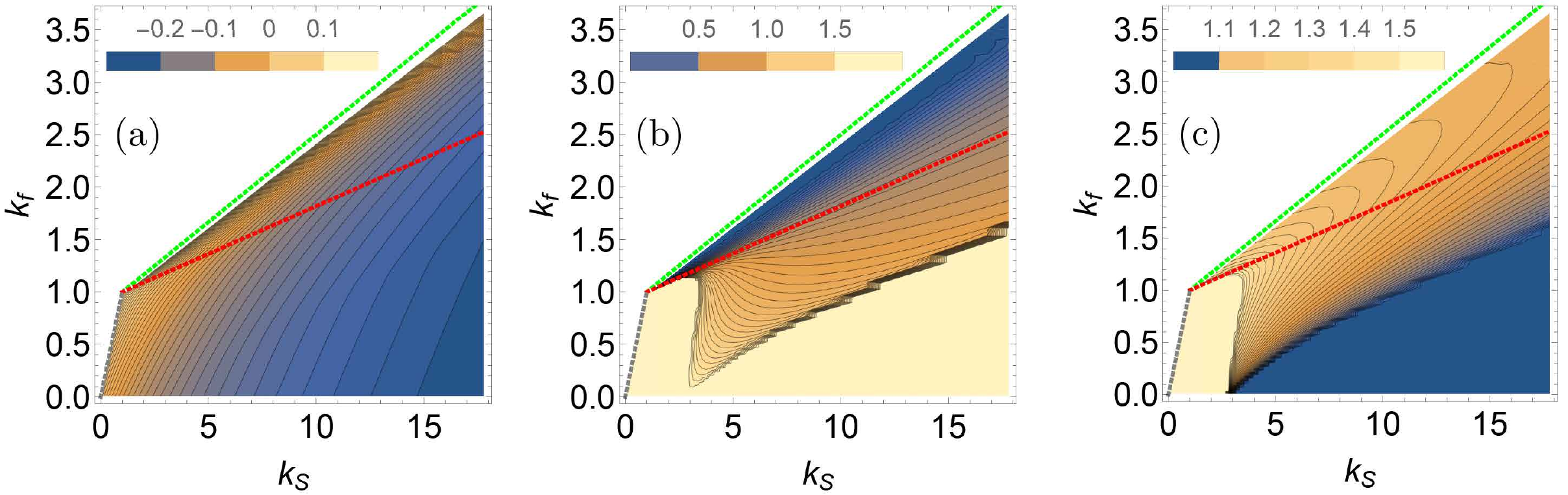}
\caption{Numerically-obtained entropy-gain minimization results for the pure-loss channel with $N_S=0.1$ and $\eta_B=0.2$.  (a) Minimum entropy gain, $E_{\Psi^c\otimes \mathcal{I}}^\star(\kappa_S,\kappa_f)$.  (b) Optimum $\delta$ value.  (c) Optimum $\gamma$ value.  In (b) and (c) the green line, $\kappa_f = (1+2\kappa_SN_S)/(1+2N_S)$, and the gray line, $\kappa_f = \kappa_S$, mark the (\ref{kf_exist}) upper limit on possible $\kappa_f$ values, and the red line, $\kappa_f = (1+\kappa_SN_S)/(1+N_S)$, is the $\kappa_f$ value below which the local minimum at $\gamma = \delta = \pi/2$ is also the global minimum for sufficiently small $\kappa_S$.}
\label{minimization_Loss}
\end{figure*}

\begin{figure*}
\includegraphics[width=0.98\textwidth]{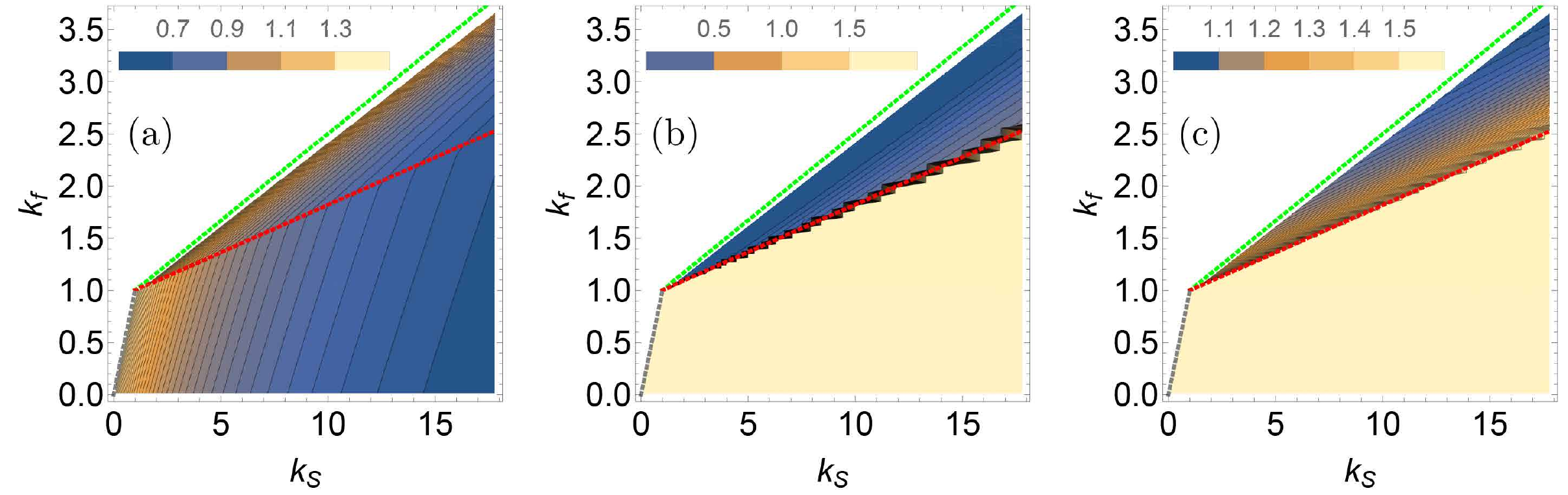}\
\caption{Numerically-obtained entropy-gain minimization results for the complementary quantum-limited amplifier channel with $N_S=0.1$ and $G_B=1.5$.  (a) Minimum entropy gain, $E_{\Psi^c\otimes \mathcal{I}}^\star(\kappa_S,\kappa_f)$.  (b) Optimum $\delta$ value.  (c) Optimum $\gamma$ value.  In (b) and (c) the green line, $\kappa_f = (1+2\kappa_SN_S)/(1+2N_S)$, and the gray line, $\kappa_f = \kappa_S$, mark the (\ref{kf_exist}) upper limit on possible $\kappa_f$ values, and the red line, $\kappa_f = (1+\kappa_SN_S)/(1+N_S)$, is the $\kappa_f$ value below which the local minimum at $\gamma = \delta = \pi/2$ is also the global minimum.}
\label{minimization_Contra}
\end{figure*}

\subsubsection{Minimization over $\gamma,\delta,\theta_v,\theta_{uv}$}
With  ${\boldsymbol \Lambda}_{SW}$ and ${\boldsymbol \Lambda}_{N'W}$ in hand, it is straightforward to obtain the symplectic eigenvalues $\nu_\pm$ and $\mu_\pm$, from which we get $E_{\Psi^c\otimes\mathcal{I}}(\rho_{SW})$ using Eq.~(\ref{munu_exp}). The only parameters to be optimized over in minimizing $E_{\Psi^c\otimes\mathcal{I}}(\rho_{SW})$ are then $\gamma,\delta,\theta_v$, $\theta_{uv}$, because the $\kappa_S$ and $\kappa_f$ constraints and are implicit in Eqs.~(\ref{para_all}). At this point it is convenient to make two further parameter changes.  First, we introduce $\zeta$ such that $\cos(\gamma)=\sqrt{(\kappa_S-\kappa_f)N_S/\kappa_f}\,\cos(\zeta)$ with $\cos^2(\zeta)\le \kappa_f/(\kappa_S-\kappa_f)N_S$, and second, we define $\xi = \theta_v + \theta_{uv}$. Then, because the $\{\nu_\pm,\mu_\pm\}$ only depend on $\gamma, \delta,$ and $\theta_v+\theta_{uv}$, we have reduced our task to minimizing a closed-form $E_{\Psi^c\otimes\mathcal{I}}(\rho_{SW})$ expression over the choice of three parameters: $\zeta\in[0,\pi/2], \delta\in[0,\pi/2],$ and $\xi \in[-\pi,\pi]$. 
 
For $\mathcal{A}_{G_B}^{0}$, $\mathcal{L}_{\eta_B}^{0}$, and $\tilde{\mathcal{A}}_{G_B}^{0}$ we find that the only solution to $\partial_{\zeta}E_{\Psi^c\otimes\mathcal{I}}(\rho_{SW})=\partial_{\delta}E_{\Psi^c\otimes\mathcal{I}}(\rho_{SW})=\partial_{\xi}E_{\Psi^c\otimes\mathcal{I}}(\rho_{SW})=0$ is $\zeta=\delta=\pi/2$ (corresponding to $\gamma=\delta=\pi/2$), at which point $\xi=\theta_v+\theta_{uv}$ can be arbitrary. For $\kappa_f\le (1+\kappa_SN_S)/(1+N_S)$, one can verify numerically that $\gamma=\delta=\pi/2$ is indeed the global minimum of $E_{\Psi^c\otimes\mathcal{I}}(\rho_{SW})$ for the $\mathcal{A}_{G_B}^{0}$ and $\tilde{\mathcal{A}}_{G_B}^{0}$ channels. The situation is more complicated for the $\mathcal{L}_{\eta_B}^{0}$ channel, because for this channel there is a parameter region in which the global minimum is not achieved at the stationary point (local minimum).  However, the convexity of $E_{\Psi^c\otimes\mathcal{I}}(\rho_{SW})$ with respect to $\kappa_S$ and $\kappa_f$---see Fig.~\ref{concavity_fig}, below---combined with Alice and Bob's choosing $\kappa_S\le 1$ for QKD, leads to the pure-loss channel's minimum $E_{\Psi^c\otimes\mathcal{I}}(\rho_{SW})$ being at its stationary point. 

Figures~\ref{minimization_Amp}--\ref{minimization_Contra} present numerically-obtained $E_{\Psi^c\otimes\mathcal{I}}(\rho_{SW})$ minimization results for the $\mathcal{A}_{G_B}^{0}$ channel (with $N_S = 0.1$ and $G_B = 1.5$), the $\mathcal{L}_{\eta_B}^{0}$ channel (with $N_S = 0.1$ and $\eta_B = 0.2$), and $\tilde{\mathcal{A}}_{G_B}^{0}$ channel (with $N_S = 0.1$ and $G_B = 1.5$), respectively.   Plotted versus $\kappa_S$ and $\kappa_f$ in each figure are:  (a) $E_{\Psi^c\otimes \mathcal{I}}^\star(\kappa_S,\kappa_f) = \min_{\rho_{SW}} E_{\Psi^c\otimes\mathcal{I}}(\rho_{SW})$, (b) the optimum $\delta$ value, and (c) the optimum $\gamma$ value.  The green line, $\kappa_f = (1+2\kappa_SN_S)/(1+2N_S)$, and the gray line, $\kappa_f = \kappa_S$, in (b) and (c) mark the (\ref{kf_exist}) upper limit on possible $\kappa_f$ values.  The red line, $\kappa_f = (1+\kappa_SN_S)/(1+N_S)$, in (b) and (c) is the $\kappa_f$ value below which the local minimum at $\gamma = \delta = \pi/2$ is also the global minimum for the amplifier channels, and for the pure-loss channel when $\kappa_S$ is sufficiently small (a region that includes $\kappa_S \le 1$, as noted earlier).  

Although Figs.~\ref{minimization_Amp}--\ref{minimization_Contra} only provide information about one set of $N_S,G_B$ and $\eta_B$ values, the behaviors shown in these figures are generic.  Indeed, we have verified that this is for $G_B = 10, 100$, and, by asymptotic expansions, for $G_B \gg 1$ and $N_S \ll 1$.  Furthermore, the asymptotic results allow us to show that the beam-splitter active injection attack achieves $E_{\Psi^c\otimes \mathcal{I}}^\star(\kappa_S,\kappa_f)$ when $G_B \gg 1$ and $N_S \ll 1$.  

\subsubsection{Asymptotic results}
The closed-form expression for $E_{\Psi^c\otimes\mathcal{I}}(\rho_{SW})$ as a function of $\zeta,\delta$ and $\xi$ is complicated, preventing us from minimizing it analytically. That is not the case, however, when $\kappa_f\simeq \kappa_S$. Physically, this corresponds to Alice and Bob's security testing confining Eve's attack to the low-intrusion regime, e.g., when Eve limits herself to a passive attack in which she only interacts with light that is lost in propagation between Alice and Bob and between Bob and Alice.   For this low-intrusion regime let us write $\kappa_f$ as 
\be
\kappa_f=(1-f_E)\kappa_S,
\label{fE}
\ee
where $0\le f_E \ll 1$ is a function of the attack parameters $\zeta, \delta$ and $\xi$, and then evaluate $E_{\Psi^c\otimes\mathcal{I}}(\rho_{SW})$ to first order in $f_E$, viz., 
\begin{eqnarray}
\lefteqn{E_{\Psi^c\otimes \mathcal{I}}(\rho_{SW})=\left.E_{\Psi^c\otimes \mathcal{I}}(\rho_{SW})\right|_{f_E = 0}
} \nonumber \\
&&+(\partial_{f_E}E_{\Psi^c\otimes \mathcal{I}}(\rho_{SW})|_{f_E=0})f_E+O(f_E^2).
\label{munu_exp2}
\end{eqnarray} 
It turns out that the zeroth-order term is independent of $\zeta, \delta$, and $\xi$.   Thus, 
Eve's optimum $\zeta,\delta$ and $\xi$ values when $0\le f_E \ll 1$ are given by 
\be
\mbox{arg min}_{\zeta,\delta,\xi}\partial_{f_E}E_{\Psi^c\otimes \mathcal{I}}(\rho_{SW})|_{f_E=0},
\label{goal_new}
\ee
and using those values in Eq.~(\ref{munu_exp2}) will then yield $E_{\Psi^c\otimes \mathcal{I}}^\star(\kappa_S,\kappa_f)$ to first order in $f_E$.
  
Note that from (\ref{kf_exist}) and (\ref{fE}), we have that
$
\kappa_S\le 1/[1-(1+2N_S)f_E],
$
whence
\ba
\mu_- |_{f_E=0}&=&1,\\
\mu_+|_{f_E=0}&=&1+2(1-\kappa_S)N_S>1.
\ea
So, to complete our asymptotic analysis, we need only find the symplectic eigenvalues, $\nu_\pm|_{f_E = 0}$, for Bob's three possible channels.  

\begin{enumerate}
\item
For pure-loss channel, $\mathcal{L}_{\eta_B}^{0}$, we find that 
\ba
\nu_- |_{f_E=0}&=&1,\\
\nu_+ |_{f_E=0}&=&1+2[1-\kappa_S(1-\eta_B)]N_S>1.
\ea
Applying $\lim_{x\to0}\partial_xg(x)= \infty$ to Eq.~(\ref{munu_exp}), we see that it suffices to consider
\be
{\rm min}_{\zeta,\delta,\xi}
\partial_{f_E}(\nu_--\mu_-)|_{f_E=0},
\label{easy_loss}
\ee
to obtain $E_{\Psi^c\otimes \mathcal{I}}^\star(\kappa_S,\kappa_f)$, 
The minimization in~(\ref{easy_loss}) can be done analytically, giving the result $\zeta=\delta=\pi/2$.

\item
For $G_B>1$, both quantum-limited amplifier,  $\mathcal{A}_{G_B}^{0}$, and its complementary channel, $\tilde{\mathcal{A}}_{G_B}^{0}$, have $\nu_+ |_{f_E=0}>\nu_- |_{f_E=0}>1$. Thus to obtain $E_{\Psi^c\otimes \mathcal{I}}^\star(\kappa_S,\kappa_f)$, it suffices to consider
\be
{\rm min}_{\zeta,\delta,\xi}
\partial_{f_E}(-\mu_-)|_{f_E=0}.
\label{easy_amp}
\ee
The minimization in~(\ref{easy_amp}) can be done analytically, giving the result $\zeta=\delta=\pi/2$.

\end{enumerate}

\begin{figure*}
\includegraphics[width=0.98\textwidth]{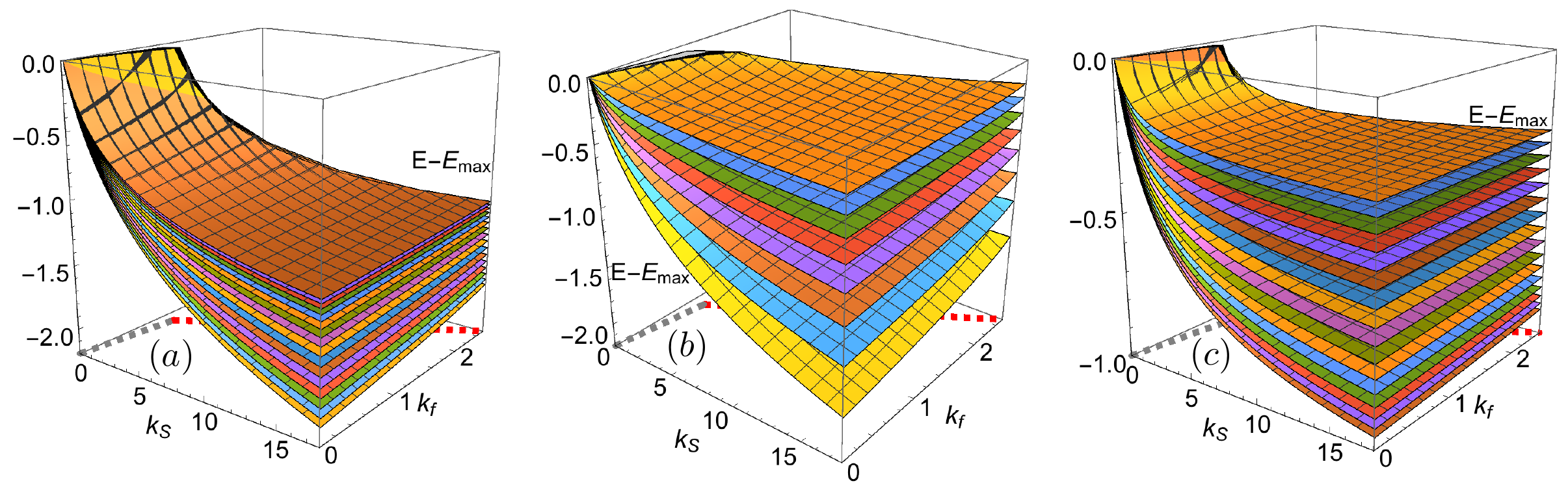}
\caption{Plots of $E - E_{\rm max}$ versus $\kappa_S$ and $\kappa_f$ for $N_S = 0.1$, with $E \equiv 
E_{\Psi^c\otimes \mathcal{I}}^\star(\kappa_S,\kappa_f)$ and $E_{\rm max} \equiv \max_{\kappa_S,\kappa_f}E_{\Psi^c\otimes \mathcal{I}}^\star(\kappa_S,\kappa_f)$. (a) Quantum-limited amplifier channel with, from bottom to top, $\log_{10}(G_B-1)$ increasing from $-1$ to 0.5 in 0.1 increments. (b) Pure-loss channel with, from top to bottom, $\eta_B$ increasing from 0.2 to 1 in 0.1 increments. (c) Contravariant quantum-limited amplifier channel with, from bottom to top, $\log_{10}(G_B-1)$ increasing from $-1$ to 0.5 in 0.1 increments.  In (a)--(c), the gray line, $\kappa_f = \kappa_S$, marks part of the (\ref{kf_exist}) upper limit on possible $\kappa_f$ values, and the red line, $\kappa_f = (1+\kappa_SN_S)/(1+N_S)$, is the $\kappa_f$ value below which the entropy-gain's local minimum at $\gamma = \delta = \pi/2$ is also its global minimum.
\label{concavity_fig}}
\end{figure*}
\subsubsection{Optimum attack}
At $\zeta=\delta=\pi/2$, we have
\begin{subequations}
\begin{align}
u_0 &= \sqrt{\kappa_f}, \label{u0opt}\\[.05in]
v_0 &= 0, \label{v0opt}\\[.05in]
{\bf u}^\dagger{\bf u} & =  (\kappa_S-\kappa_f)N_S + 1 - \kappa_f,\label{u0normOpt}\\[.05in]
{\bf v}^\dagger{\bf v} &= (\kappa_S-\kappa_f)N_S, \label{v0normOpt}\\[.05in]
{\bf v}^\dagger{\bf u} &= 0 \label{vuOpt},
\end{align}
\label{beamsplitter_attack}
\end{subequations}
which are the parameter values of the beam-splitter injection attack considered in the Ref.~\cite{Quntao_2015}.  With the optimum parameters given by Eqs.~(\ref{beamsplitter_attack}), we can evaluate $E_{\Psi^c\otimes \mathcal{I}}^\star(\kappa_S,\kappa_f)$ via Eq.~(\ref{munu_exp}). Combined with $\max_{\rho_{SW}} = S(\rho_B)$, we obtain $\chi_E(\kappa_S,\kappa_f)$ in Eq.~(\ref{final_opt_main}).

In particular, the covariance matrix, ${\boldsymbol \Lambda}^\star_{SW}$, of the optimum input state, $\rho^\star_{SW}$, is  
\begin{equation}
{\boldsymbol \Lambda}_{SW}^\star =  \frac{1}{4}\left[\begin{array}{cc}
(1+2\kappa_S N_S){\bf I}_2 & \sqrt{k_f}\,{\bf C}_{YW} \\
\sqrt{k_f}{\bf C}_{YW} & {\bf A}_W \end{array}\right],
\label{RW_opt}
\end{equation} 
with symplectic eigenvalues $\mu_\pm^\star$.
The optimum output state, $\rho^\star_{N'W}$ and its covariance matrix, ${\boldsymbol \Lambda}^\star_{N'W}$ depend on which channel Bob employs.  
\begin{enumerate}
\item For the $\mathcal{A}_{G_B}^{0}$ channel, we get
\begin{align}
&{\boldsymbol \Lambda}^\star_{N'W} =
\nonumber
\\ &\frac{1}{4}\!\left[\begin{array}{ccc} {\bf A}_{N'}^\star & & \sqrt{k_f(G_B-1)}\,{\bf C}_{YW} \\[.05in]
\sqrt{k_f(G_B-1)}\,{\bf C}_{YW}  & & \end{array}\right],
\end{align}
with ${\bf A}_{N'}^\star = [1+ 2(G_B-1)(1+\kappa_SN_S)]{\bf I}_2$.
\item For the $\mathcal{L}_{\eta_B}^{0}$ channel, we get
\begin{align}
&{\boldsymbol \Lambda}^\star_{N'W} =
\nonumber
\\
&\frac{1}{4}\!\left[\begin{array}{ccc} [1+ 2(1-\eta_B)\kappa_SN_S]{\bf I}_2 & & \sqrt{k_f(1-\eta_B)\,}{\bf C}_{YW}\\[.05in]
\sqrt{k_f(1-\eta_B)}\,{\bf C}_{YW} & & {\bf A}_W \end{array}\right].
\end{align} 

\item For the $\tilde{\mathcal{A}}_{G_B}^{0}$ channel, we get
\begin{align}
&{\boldsymbol \Lambda}^\star_{N'W} =
\nonumber
\\
&\frac{1}{4}\!\left[\begin{array}{ccc} -1+2G_B[1+ 2(1+\kappa_SN_S)]{\bf I}_2 & & \sqrt{G_Bk_f}\,{\bf C}_{YW}\\[.05in]
\sqrt{G_Bk_f}\,{\bf C}_{YW} & & {\bf A}_W \end{array}\right].
\end{align} 
\end{enumerate}
With $\nu_\pm^\star$ denoting the symplectic eigenvalues of ${\boldsymbol \Lambda}^\star_{N'W} $, we have that ,
\be
\chi_E(\kappa_S,\kappa_f)=g(N_B)-
E_{\Psi^c\otimes \mathcal{I}}^\star(\kappa_S,\kappa_f),
\ee
with
\begin{align}
E_{\Psi^c\otimes \mathcal{I}}^\star(\kappa_S,\kappa_f)&=g[(4\nu^\star_+-1)/2]+g[(4\nu^\star_--1)/2]
\nonumber
\\
&
-g[(4\mu^\star_+-1)/2]-g[(4\mu^\star_--1)/2].
\label{final_opt}
\end{align}

\end{proof}

We complete this section by presenting our numerical verification, shown in Fig.~\ref{concavity_fig}, that $E_{\Psi^c\otimes \mathcal{I}}^\star(\kappa_S,\kappa_f)$ is convex, where, for better visualization, we have plotted $E - E_{\rm max}$ with $E \equiv 
E_{\Psi^c\otimes \mathcal{I}}^\star(\kappa_S,\kappa_f)$ and $E_{\rm max} \equiv \max_{\kappa_S,\kappa_f}E_{\Psi^c\otimes \mathcal{I}}^\star(\kappa_S,\kappa_f)$.  Although these plots assume $N_S = 0.1$, we have verified that similar behaviors prevail at other $N_S$ values of interest.  

\section{Proof of Theorem~\ref{theorem_opt_attack_general} \label{AppB}}

Our proof uses the fact that performing arbitrary local unitaries on the ${\bm W}$ modes preserves $F(\rho_{S_m{\bm W}})$. In particular, we have the following lemma; see Appendix~\ref{AppC} for its proof.
\begin{lemma}
For the Eq.~(\ref{Q2_general}) constraint, i.e.,
\be
\sum_{n=1}^M\left[|\braket{\hat{a}_{S_m}\hat{a}_{W_n}}|^2+|\braket{\hat{a}^\dagger_{S_m}\hat{a}_{W_n}}|^2\right]=K_f^{(m)} C_S^2,
\label{Q2_general_proof}
\ee 
we can apply beam-splitter unitaries to the ${\bm W}$ modes that result in output modes $\tilde{\bm W}$ modes that reduce Eq.~(\ref{Q2_general_proof}) to  
\be
{|\braket{\hat{a}_{S_m}\hat{a}_{\tilde{W}_1}}|^2+|\braket{\hat{a}_{S_m}^\dagger\hat{a}_{\tilde{W}_1}}|^2+|\braket{\hat{a}_{S_m}\hat{a}_{\tilde{W}_2}}|^2}=K_f^{(m)} C_S^2.
\label{non_zeros}
\ee 
\label{lemma_form1}
\end{lemma}
In what follows we shall omit the tildes on the preceding output modes.   Bear in mind that the beam-splitter unitaries that we are using for this proof are a \emph{conceptual} tool, i.e., they do \emph{not} need to be implemented in the TW-QKD system.  

With Lemma~\ref{lemma_form1} in hand, we will maximize $F(\rho_{S_m{\bm W}})$ over the reduced density operator $\rho_{S,W_1,W_2}$ subject to the constraints from Eq.~(\ref{Q1}),
\be
\braket{\hat{a}^\dagger_{S_m}\hat{a}_{S_m}}=\kappa_S^{(m)}N_S,
\label{Q1_proof_2},
\ee
and~(\ref{non_zeros}), to obtain
\be
\chi_{E}^\prime\!\left(\kappa_S^{(m)},K_f^{(m)}\right)=\max_{\rho_{S_mW_1W_2}}[ 
S(\rho_B)-E_{\Psi^c\otimes \mathcal{I}}(\rho_{S_mW_1W_2})].
\label{goal_proof_2}
\ee
It can be shown that the ${\bm W}$ modes which emerge from Lemma~\ref{lemma_form1}'s beam-splitter unitaries are still in independent, identically-distributed thermal states with average photon number $N_S$.  Also, Ref.~\cite{Zhuang_2016_cl}'s subadditivity result implies we need only  consider $\rho_{S_mW_1W_2}$ that are Gaussian. 
Hence our goal for completing Theorem~\ref{theorem_opt_attack_general}'s proof is maximizing Eq.~(\ref{goal_proof_2}) for Gaussian $\rho_{S_mW_1W_2}$ that obey Eqs.~(\ref{non_zeros}) and~(\ref{Q1_proof_2}).  In the rest of the proof, which is similar to what we did in Appendix~\ref{AppA}, we will omit the $m$ subscripts and superscripts and use $S, W_1,W_2$ to denote the three modes under consideration.

To begin, we note that the $S(\rho_B)$ maximization from Appendix~\ref{entropy_first} applies in the present circumstances, i.e., $\max_{\rho_{SW_1W_2}}S(\rho_B) = g(N_B)$, and this maximum is achieved by having $\braket{\hat{a}_S} = 0$.  
The optimum Gaussian state $\rho_{S W_1W_2}$ is therefore zero-mean with $\langle\hat{a}_S^\dagger\hat{a}_S\rangle = \kappa_SN_S$, $\langle \hat{a}_{W_1}^\dagger\hat{a}_{W_1}\rangle = \langle \hat{a}_{W_2}^\dagger\hat{a}_{W_2}\rangle = N_S$, and $\langle \hat{a}_{W_1}^\dagger\hat{a}_{W_2}\rangle = \langle\hat{a}_{W_1}\hat{a}_{W_2}\rangle = 0$, so four additional complex-valued parameters, $\braket{\hat{a}_S^2}$,
$\braket{\hat{a}_S\hat{a}_{W_1}}$,
$\braket{\hat{a}_S^\dagger\hat{a}_{W_1}}$, 
$\braket{\hat{a}_S\hat{a}_{W_2}}$---equivalently eight real parameters---complete its characterization. 

Now, by appropriate phase shifts of the $S$, $W_1$ and $W_2$ modes---which will not affect the entropy-gain term---we can assume that
\ba
\braket{\hat{a}_S^2}&=&c_1\ge0, \mbox{ where $c_1\le \kappa_SN_S$},\\[.05in]
\braket{\hat{a}_S\hat{a}_{W_1}}&=&a_1\ge 0,\\[.05in]
\braket{\hat{a}_S^\dagger\hat{a}_{W_1}}&=&b_1 e^{i \theta}, \mbox{ where $b_1\ge 0, \theta\in[0,2\pi)$},\\[.05in]
\braket{\hat{a}_S\hat{a}_{W_2}}&=&a_2\ge 0.
\ea
Consequently, the entropy-gain minimization,
\ba
&\min_{\rho_{SW_1W_2}} 
E_{\Psi^c\otimes \mathcal{I}}(\rho_{SW_1W_2})
= \nonumber\\[.05in]
&\min_{\rho_{SW_1W_2}}[S(\rho_{N'W_1W_2})-S(\rho_{SW_1W_2})],
\label{goal_proof_En_2}
\ea
will only involve five parameters, $\{c_1,a_1,b_1,\theta,a_2\}$, of which only four are independent, because Eq.~(\ref{non_zeros}) implies that  
\be
a_1^2+b_1^2+a_2^2 = K_f C_S^2.
\label{constraint_2}
\ee
Furthermore, with $\{\nu_k : 1\le k\le 3\}$ and $\{\mu_k : 1\le k \le 3\}$ being the symplectic eigenvalues of the covariance matrices ${\boldsymbol \Lambda}_{SW_1W_2}$ and ${\boldsymbol \Lambda}_{N'W_1W_2}$, we have that
\begin{eqnarray}
\lefteqn{E_{\Psi^c\otimes \mathcal{I}}(\rho_{SW_1W_2})= } \nonumber \\[.05in]
&& \sum_{k=1}^3\{g[(4\nu_k-1)/2]-g[(4\mu_k-1)/2]\}.
\label{munu_exp_2}
\end{eqnarray}

The covariance matrices that we need are given as follows.  For ${\boldsymbol \Lambda}_{SW_1W_2}$ we have that
\begin{equation}
{\boldsymbol \Lambda}_{SW_1W_2} = \frac{1}{4}\!\left[\begin{array}{ccc}  
{\bf A}_S& {\bf C}_{SW_1}& {\bf C}_{SW_2} 
\\[.05in]
{\bf C}_{SW_1}& {\bf A}_W& {\bf 0}
\\[.05in]
{\bf C}_{SW_2} & {\bf 0}&  {\bf A}_W
\end{array}\right],
\end{equation}
where 
\begin{equation}
{\bf A}_{S} =  \left[\begin{array}{cc} 1+2(\kappa_SN_S+c_1)&0
\\[.05in]
0& 1+2(\kappa_SN_S-c_1)
\end{array}\right],
\end{equation}
\begin{equation}
{\bf C}_{SW_1} = 2\left[\begin{array}{ccc} 
a_1+b_1\cos\theta & & b_1\sin\theta
\\[.05in]
b_1\sin\theta & & -a_1+b_1\cos\theta\end{array}\right],
\end{equation}
${\bf C}_{SW_2}=2a_2\,{\rm Diag}[1,-1]$, and ${\bf A}_W = (2N_S+1){\bf I}_2$.  
For ${\boldsymbol \Lambda}_{N'W_1W_2}$, however, we need expressions for each of Bob's three channels. 
\begin{enumerate}
\item For the $\mathcal{A}^0_{G_B}$ channel, we get
\begin{equation}
{\boldsymbol \Lambda}_{N^\prime W_1W_2} = \frac{1}{4}\!\left[\begin{array}{ccc}  
{\bf A}_{N'}& {\bf C}_{N'W_1}& {\bf C}_{N'W_2} 
\\[.05in]
{\bf C}_{N'W_1}& {\bf A}_W& {\bf 0}
\\[.05in]
{\bf C}_{N'W_2} & {\bf 0}&  {\bf A}_W
\end{array}\right],
\end{equation}
where
\begin{equation}
{\bf A}_{N'} =  \left[\begin{array}{cc} 1+2x_{N'+}&0
\\[.05in]
0& 1+2x_{N'-}
\end{array}\right],
\end{equation} 
with $x_{N'\pm}=(G_B-1)(1+\kappa_SN_S\pm c_1)$,
\begin{equation}
{\bf C}_{N'W_1} = 2\sqrt{G_B-1}\left[\begin{array}{ccc} 
a_1+b_1\cos\theta & & b_1\sin\theta
\\[.05in]
-b_1\sin\theta & & a_1-b_1\cos\theta\end{array}\right],
\end{equation}
and ${\bf C}_{N'W_2}=2\sqrt{G_B-1}\,a_2\,{\rm Diag}[1,1]$.

\item  For the $\mathcal{L}^0_{\eta_B}$ channel, we get 
\begin{equation}
{\boldsymbol \Lambda}_{N^\prime W_1W_2} = \frac{1}{4}\!\left[\begin{array}{ccc}  
{\bf A}_{N'}& {\bf C}_{N'W_1}& {\bf C}_{N'W_2} 
\\[.05in]
{\bf C}_{N'W_1}& {\bf A}_W& {\bf 0}
\\[.05in]
{\bf C}_{N'W_2} & {\bf 0}&  {\bf A}_W
\end{array}\right],
\end{equation}
where 
\begin{equation}
{\bf A}_{N'} =  \left[\begin{array}{cc} 1+2x_{N'+}&0
\\[.05in]
0& 1+2x_{N'-}
\end{array}\right],
\end{equation}
with $x_{N\pm}=(1-\eta_B)(\kappa_SN_S\pm c_1)$,
${\bf C}_{N'W_1} = \sqrt{1-\eta_B}\,{\bf C}_{SW_1}$, and  ${\bf C}_{N'W_2}=\sqrt{1-\eta_B}\,{\bf C}_{SW_2}$ 

\item  For the $\tilde{\mathcal{A}}^0_{G_B}$ channel, we get
\begin{equation}
{\boldsymbol \Lambda}_{N^\prime W_1W_2} = \frac{1}{4}\!\left[\begin{array}{ccc}  
{\bf A}_{N'}& {\bf C}_{N'W_1}& {\bf C}_{N'W_2} 
\\[.05in]
{\bf C}_{N'W_1}& {\bf A}_W& {\bf 0}
\\[.05in]
{\bf C}_{N'W_2} & {\bf 0}&  {\bf A}_W
\end{array}\right],
\end{equation} 
where
\begin{equation}
{\bf A}_{N'} =  \left[\begin{array}{cc} -1+2x_{N'+}&0
\\[.05in]
0& -1+2x_{N'-}
\end{array}\right],
\end{equation} 
with $x_{N'\pm}=G_B(1+\kappa_SN_S\pm c_1)$, 
${\bf C}_{N'W_1} = \sqrt{G_B}\,{\bf C}_{SW_1}$, and ${\bf C}_{N'W_2}=\sqrt{G_B}\,{\bf C}_{SW_2}$
\end{enumerate}

At this point it is possible---for all three of Bob's channels---obtain closed-form expressions for the entropy gain that are functions of $\{c_1,a_1,b_1,\theta,a_2\}$.  In principle, these expressions can be minimized, subject to Eq.~(\ref{constraint_2}), but in practice they are too complicated for that to be done analytically.  Numerical minimization can be done, however, for which transforming to
\begin{align}
&c_1=\kappa_S N_S \cos^2(\tau_r),
\\[.05in]
&a_1=\sqrt{K_f }\,C_S\cos(\tau_1),
\\[.05in]
&b_1=\sqrt{K_f }\,C_S\sin(t_1)\cos(\tau_2), 
\\[.05in]
&a_2=\sqrt{K_f }\,C_S\sin(t_1)\sin(\tau_2),
\end{align}
with $\tau_r\in [0,\pi/2]$, $\tau_1 \in [0,\pi/2]$, and $\tau_2 \in [0,\pi]$, automatically ensures that Eq.~(\ref{constraint_2}) is satisfied, and reduces the entropy gain's numerical minimization to a four-dimensional optimization.  

The preceding analysis completes the proof of Theorem~\ref{theorem_opt_attack_general}  modulo our proving Lemma~\ref{lemma_form1}, which we accomplish in Appendix~\ref{AppC}.  

\section{Proof of Lemma~\ref{lemma_form1}\label{AppC}}
\begin{proof} 
Our objective is to show that a collection of beam-splitter unitaries involving the ${\bm W}$ modes can reduce Eq.~(\ref{Q2_general_proof}) to Eq.~(\ref{non_zeros}).  We begin by showing how to eliminate the undesired phase-insensitive cross correlations, i.e., the $\braket{\hat{a}_{S_m}^\dagger\hat{a}_{W_n}}\}$ for $2\le n \le M$.  First, we apply phase shifts to the ${\bm W}$ modes so that all $\braket{\hat{a}_{S_m}^\dagger\hat{a}_{W_n}}\} \ge 0$ for $1\le n \le M$, with $\braket{\hat{a}_{S_m}^\dagger\hat{a}_{W_1}} >0$~\cite{footnote8}.  Next, starting with $n=2$ and continuing until $n = M$, we use beam splitters to effect the following transformations,
\ba
\hat{a}_{W_1}^{(n)}&=&\sqrt{1-\eta_n}\,\hat{a}_{W_1}^{(n-1)}+\sqrt{\eta_n}\,\hat{a}_{W_n},
\nonumber
\\
\hat{a}_{W_n}^\prime&=&\sqrt{\eta_n}\,\hat{a}_{W_1}^{(n-1)}-\sqrt{1-\eta_n}\, \hat{a}_{W_n},
\ea 
with 
\be
\eta_n\equiv\frac{\braket{\hat{a}_{S_m}^\dagger\hat{a}_{W_n}}^2}{\braket{\hat{a}_{S_m}^\dagger\hat{a}_{W_1}^{(n-1)}}^2+ \braket{\hat{a}_{S_m}^\dagger\hat{a}_{W_n}}^2},
\ee
where $\hat{a}_{W_1}^{(1)} \equiv \hat{a}_{W_1}$ is the $W_1$ mode's initial photon-annihilation operator, and $\hat{a}_{W_n}'$, for $2\le n \le M$, is the $W_n$ mode's photon annihilation operator \emph{after} its beam-splitter transformation.  For $2\le n \le M$, it is easily verified that this process results in
\ba
\braket{\hat{a}_{S_m}^\dagger\hat{a}_{W_1}^{(n)}}&=&\sqrt{\braket{\hat{a}_{S_m}^\dagger\hat{a}_{W_1}^{(n-1)}}^2+ \braket{\hat{a}_{S_m}^\dagger\hat{a}_{W_n}}^2},
\nonumber
\\[.05in]
\braket{\hat{a}_{S_m}^\dagger\hat{a}_{W_n}'}&=&0.
\ea
Collapsing this iteration into a single formula gives us our desired result,
\ba
\braket{\hat{a}_{S_m}^\dagger\hat{a}_{W_1}^{(M)}}&=&\sqrt{{\sum_{n=1}^M \braket{\hat{a}_{S_m}^\dagger\hat{a}_{W_n}}^2}},
\nonumber
\\[.05in]
\braket{\hat{a}_{S_m}^\dagger\hat{a}_{W_n}^{\prime}}&=&0,\ \mbox{ for $2\le n\le M$}.
\label{O1_output}
\ea
It is also straightforward to obtain expressions for $\braket{\hat{a}_{S_m}\hat{a}_{W_1}^{(M)}}$ and $\{\braket{\hat{a}_{S_m}\hat{a}'_{W_n}} : 2\le n \le M\}$, all of which, in general, will be nonzero.  To suppress the unwanted phase-sensitive cross correlations, we parallel what we just did for the phase-insensitive case.

First, we apply phase shifts to $\{\hat{a}'_{W_n} : 2 \le n \le M\}$ so that all $\braket{\hat{a}_{S_m}\hat{a}'_{W_n}} \ge 0$ for $2\le n \le M$, with $\braket{\hat{a}_{S_m}\hat{a}'_{W_2}} > 0$~\cite{footnote7}.  Next, starting with $n=3$ and continuing until $n=M$, we use beam splitters to effect the following transformations,
\ba
\hat{a}_{W_2}^{\prime (n)}&=&\sqrt{1-\eta'_n}\,\hat{a}_{W_2}^{\prime (n-1)}+\sqrt{\eta_n}\,\hat{a}'_{W_n},
\nonumber
\\
\hat{a}^{\prime\prime}_{W_n}&=&\sqrt{\eta'_n}\,\hat{a}_{W_2}^{\prime (n-1)}-\sqrt{1-\eta'_n}\, \hat{a}'_{W_n},
\ea 
with 
\be
\eta'_n\equiv\frac{\braket{\hat{a}_{S_m}\hat{a}'_{W_n}}^2}{\braket{\hat{a}_{S_m}\hat{a}_{W_2}^{\prime (n-1)}}^2+ \braket{\hat{a}_{S_m}\hat{a}'_{W_n}}^2},
\ee
where $\hat{a}_{W_2}^{\prime (2)} \equiv \hat{a}'_{W_2}$ is the $W'_2$ mode's initial photon-annihilation operator, and $\hat{a}''_{W_n}$, for $3\le n \le M$, is the $W'_n$ mode's photon annihilation operator \emph{after} its beam-splitter transformation.  For $3\le n \le M$, it is easily verified that this process results in
\ba
\braket{\hat{a}_{S_m}\hat{a}_{W_2}^{\prime (n)}}&=&\sqrt{\braket{\hat{a}_{S_m}\hat{a}_{W_2}^{\prime (n-1)}}^2+ \braket{\hat{a}_{S_m}\hat{a}'_{W_n}}^2},
\nonumber
\\[.05in]
\braket{\hat{a}_{S_m}\hat{a}''_{W_n}}&=&0.
\label{no_phase_sens}
\ea
Finally, because the beam-splitter transformations preserve total correlations, we have that 
\ba
&&\sum_{n=1}^M |\braket{\hat{a}_{S_m}\hat{a}_{W_n}}|^2 \nonumber \\[.05in]
&&=|\braket{\hat{a}_{S_m}\hat{a}_{W_1}^{(M)}}|^2  + \sum_{n=2}^M|\braket{\hat{a}_{S_m}\hat{a}'_{W_n}}|^2
\\[.05in]
 &&= |\braket{\hat{a}_{S_m}\hat{a}_{W_1}^{(M)}}|^2 + |\braket{\hat{a}_{S_m}\hat{a}_{W_2}^{\prime(M)}}|^2 \nonumber\\[.05in]
&&\hspace*{.2in}+ \sum_{n=3}^M|\braket{\hat{a}_{S_m}\hat{a}_{W_n}^{\prime\prime}}|^2
\nonumber\\
&&=|\braket{\hat{a}_{S_m}\hat{a}_{W_1}^{(M)}}|^2 + |\braket{\hat{a}_{S_m}\hat{a}_{W_2}^{\prime(M)}}|^2,
\label{penult}
\ea
where we used Eq.~(\ref{no_phase_sens}) in Eq.~(\ref{penult}). Combining Eq.~(\ref{O1_output}) and Eq.~(\ref{penult}), we complete the proof.

\end{proof}

\flushleft


\begin{thebibliography}{41}
\bibitem{footnote0}The PLOB bound, states that secret bits per mode cannot exceed $-\log_2(1-\eta)$ for a channel with transmissivity $\eta$.  It was preceded by the TGW bound~\cite{TGWbound}, which stated that secret bits per mode is bounded above by $\log_2[(1+\eta)/(1-\eta)]$ for a channel with transmissivity $\eta$.  Unlike the TGW bound, however, the PLOB bound is achievable~\cite{pirandola2017fundamental}.
\bibitem{footnote1} For simplicity, this figure presumes single-mode symbol encoding, but multi-mode symbol encoding, as employed FL-QKD, will be included in the information-gain bounds to be develped below.
\bibitem{footnote2} In FL-QKD, Alice mixes amplified spontaneous emission (ASE) into the her $Y$ mode, hence her only being able to access part of $Y$'s purification $W$; see Sec.~\ref{protocol2} for more information.
\bibitem{footnote3}Bob's number of modes per encoded symbol is $M_E = T_BW_A$, where $T_B$-s is the time duration of Bob's symbol and $W_A$-Hz is the bandwidth of Alice's transmitted light. Side-channel attacks are not considered in the current framework, i.e., we assume that the $M_E$ modes are all that enters Bob's lab in each symbol duration;As a first step to ward off Eve's attacking the QKD protocol, by illuminating Bob's terminal with out-of-band light that would be encoded by Bob and returned to her, we consider Bob employing an optical filter---not shown in Fig.~\ref{most_general}---to limit the light entering his encoder to the spectral region of Alice's transmitted light.
\bibitem{footnote4}We will assume that $cT_B$, where $c$ is light speed, is a distance that can be wholly confined within Bob's laboratory, making it impossible for Eve to employ intra-symbol feedback in her most general block-coherent attack.
\bibitem{footnote6}It is noteworthy that binary encoding, as used in FL-QKD, has more efficient reconciliation schemes than those currently available for continuous-variable encoding, as used in the TMSV protocol.
\bibitem{footnote8}There is no loss of generality in assuming that $\braket{\hat{a}_{S_m}^\dagger\hat{a}_{W_1}} \neq 0$. If all the $\braket{\hat{a}_{S_m}^\dagger\hat{a}_{W_n}}$ vanish then, no beam-splitter transformations are needed to suppress the unwanted phase-insensitive cross correlations, and, if at least one of them is nonzero, we can take that ${\bm W}$ mode and call it the $n=1$ mode.  The same will be true, below, when we null out the unwanted phase-sensitive cross correlations, i.e., there will either be a $2 \le n \le M$ value for which $\braket{\hat{a}_{S_m}\hat{a}'_{W_n}}$ is nonzero---in which case we relabel that mode as the $n=2$ mode---or there are no phase-sensitive cross correlations that must be suppressed.
\end{thebibliography}
\end{document}